\pdfoutput=1

\documentclass[conference]{IEEEtran}

\IEEEoverridecommandlockouts
\usepackage{cite}
\usepackage{amsmath,amssymb,amsfonts}
\usepackage{algorithmic}
\usepackage{graphicx}
\usepackage{textcomp}
\usepackage{xcolor}
\def\BibTeX{{\rm B\kern-.05em{\sc i\kern-.025em b}\kern-.08em
    T\kern-.1667em\lower.7ex\hbox{E}\kern-.125emX}}

\usepackage{array}
\newcolumntype{H}{>{\setbox0=\hbox\bgroup}c<{\egroup}@{}}

\usepackage{color}
\usepackage{colortbl} 
\usepackage{listings}
\usepackage{lineno}
\usepackage{hyperref}
\usepackage{url}
\usepackage{enumitem}
\usepackage{xspace}
\usepackage{enumitem}
\usepackage{balance}
\usepackage{subcaption}
\usepackage{wrapfig}
\usepackage{verbatimbox}
\usepackage{multirow}
\usepackage{multicol}
\usepackage{enumitem}
\usepackage{dblfloatfix}
\usepackage{booktabs}
\usepackage{diagbox}
\usepackage{pifont}
\usepackage[multiple]{footmisc}
\usepackage[most]{tcolorbox}
\usepackage{newfloat}
\usepackage[frozencache=true,cachedir=minted_cache]{minted}
\usepackage{adjustbox}
\usepackage{bm}
\usepackage{pdfrender}

\definecolor{WowColor}{rgb}{.75,0,.75}
\definecolor{SubtleColor}{rgb}{0,0,.50}
\newcommand{\cmark}{\textcolor{green}{\ding{51}}\xspace}%
\newcommand{\xmark}{\textcolor{red}{\ding{55}}\xspace}%

\ifdefined\Comment
        \renewcommand{\Comment}[1]{}
\else
        \newcommand{\Comment}[1]{}
\fi

\newcounter{margincounter}

\newcommand{\mypara}[1]{\vspace{.03in}\noindent \textbf{#1.}}

\definecolor{ghgreen}{rgb}{0.90,1,0.93}
\definecolor{ghred}{rgb}{1,0.88,0.94}

\definecolor{codegreen}{rgb}{0,0.6,0}
\definecolor{codegray}{rgb}{0.5,0.5,0.5}
\definecolor{codepurple}{rgb}{0.58,0,0.82}
\definecolor{backcolour}{rgb}{0.95,0.95,0.92}
\definecolor{anti-flashwhite}{rgb}{0.98, 0.98, 0.99}

\lstdefinestyle{cpp}{  
    language=C,  
    backgroundcolor=\color{anti-flashwhite},
    commentstyle=\color{codegreen},
    keywordstyle=\color{codepurple},
    numberstyle=\tiny\color{codegray},
    stringstyle=\color{blue},
    basicstyle=\setlength{\lineskip}{0pt}\footnotesize\ttfamily,
    breakatwhitespace=false,
    breaklines=true,
    captionpos=b,
    keepspaces=true,
    numbers=left,
    numbersep=5pt,
    tabsize=4,
    xleftmargin=5pt,
    columns=fullflexible,
    frame=lr,
    framerule=0pt
}

\lstdefinestyle{java}{  
    language=Java,  
    backgroundcolor=\color{anti-flashwhite},
    commentstyle=\color{codegreen},
    keywordstyle=\color{codepurple},
    numberstyle=\tiny\color{codegray},
    stringstyle=\color{blue},
    basicstyle=\setlength{\lineskip}{0pt}\footnotesize\ttfamily,
    breakatwhitespace=false,
    breaklines=true,
    captionpos=b,
    keepspaces=true,
    numbers=left,
    numbersep=5pt,
    tabsize=4,
    xleftmargin=5pt,
    columns=fullflexible,
    frame=lr,
    framerule=0pt,
    aboveskip=1.5em,
    belowskip=1em,
}

\lstdefinelanguage{MyPrompt}{
  keywords={System, Response, User},
  comment=[l]{//}, 
  morecomment=[s]{/*}{*/}
}
\lstdefinestyle{mypromptstyle}{  
    language=MyPrompt,  
    backgroundcolor=\color{backcolour},
    basicstyle=\footnotesize\ttfamily,
    commentstyle=\color{codegreen},
    keywordstyle=\color{codepurple}\bfseries,
    numberstyle=\tiny\color{codegray},
    stringstyle=\color{blue},    
    breakatwhitespace=true,
    breaklines=true,
    captionpos=b,
    keepspaces=true,
    numbers=left,
    numbersep=5pt,
    tabsize=4,
    columns=fullflexible,
    breakindent=0pt
}

\definecolor{babyblue}{rgb}{0.54, 0.81, 0.94}
\definecolor{apricot}{rgb}{0.98, 0.81, 0.69}
 \definecolor{honeydew}{rgb}{0.94, 1.0, 0.94}

\newcommand{\prediction}[1]{
\begin{tcolorbox}[
colback=pink!10, colframe=white, boxsep=1pt, arc=1pt, boxrule=1pt]
\textbf{GPT-4's response:} #1
\end{tcolorbox}
}

\newcommand{\clamasvnprediction}[1]{
\begin{tcolorbox}[
colback=pink!10, colframe=white, boxsep=1pt, arc=1pt, boxrule=1pt]
\textbf{Codellama-7B's response:} #1
\end{tcolorbox}
}

\newcommand{\clamathrprediction}[1]{
\begin{tcolorbox}[
colback=pink!10, colframe=white, boxsep=1pt, arc=1pt, boxrule=1pt]
\textbf{Codellama-13B's response:} #1
\end{tcolorbox}
}

\newcommand{\cellhl}[1]{\cellcolor{babyblue}\textbf{#1}}

\newcommand{\positive}{\cmark}
\newcommand{\negative}{\xmark }
\newcommand{\generic}{Basic\xspace}
\newcommand{\df}{CWE-DF\xspace}

\newcommand{\cwespecific}{CWE\xspace}
\newcommand{\fewshot}{CWE-Few-shot\xspace}
\newcommand{\chot}{CWE-CoT\xspace}

\newcommand{\dfstr}{Dataflow analysis-based prompt\xspace}

\newcommand{\basic}{Basic prompt\xspace}
\newcommand{\cwespecificstr}{CWE specific prompt\xspace}

\newcommand{\julietj}{Juliet Java\xspace}
\newcommand{\julietc}{Juliet C/C++\xspace}
\newcommand{\owasp}{OWASP\xspace}
\newcommand{\cvej}{CVEFixes Java\xspace}
\newcommand{\cvec}{CVEFixes C/C++\xspace}

\newcommand{\owaspabv}{OWASP\xspace}

\newcommand{\bigvul}{BigVul\xspace}

\newcommand{\clama}{CodeLlama\xspace}

\newcommand{\clamasvn}{\clama{}-7B\xspace}
\newcommand{\clamathr}{\clama{}-13B\xspace}
\newcommand{\clamathirtyfour}{\clama{}-34B\xspace}

\newcommand{\clamasvnabv}{\clamasvn}
\newcommand{\clamathrabv}{\clamathr}
\newcommand{\clamathirtyfourabv}{\clamathirtyfour}

\newcommand{\llamaseventy}{Llama-3.1-70B\xspace}
\newcommand{\llamaeight}{Llama-3.1-8B\xspace}
\newcommand{\gemini}{Gemini-1.5-Flash\xspace}
\newcommand{\qwenfourteen}{Qwen-2.5-14B\xspace}
\newcommand{\qwenthirtytwo}{Qwen-2.5-32B\xspace}
\newcommand{\qwenc}{Qwen-2.5C-1.5B\xspace}
\newcommand{\qwencseven}{Qwen-2.5C-7B\xspace}
\newcommand{\codestral}{Codestral-22B\xspace}
\newcommand{\dscoderseven}{DSCoder-7B\xspace}
\newcommand{\dscoderfifteen}{DSCoder-15B\xspace}
\newcommand{\dscoderthirtythree}{DSCoder-33B\xspace}

\begin{document}
\newcounter{mylessoncounter}
\newcounter{myreccounter}
\newcounter{rqcounter}
\setcounter{rqcounter}{1}

\title{Understanding the Effectiveness of Large Language Models in Detecting Security Vulnerabilities}


\makeatletter
\newcommand{\linebreakand}{%
  \end{@IEEEauthorhalign}
  \hfill\mbox{}\par
  \mbox{}\hfill\begin{@IEEEauthorhalign}
}
\makeatother

\author{
\IEEEauthorblockN{Avishree Khare$^*$}
\IEEEauthorblockA{
    \textit{University of Pennsylvania}\\
    Philadelphia, USA\\
    akhare@seas.upenn.edu
}
\and
\IEEEauthorblockN{Saikat Dutta$^*$}
\IEEEauthorblockA{
    \textit{Cornell University}\\
    Ithaca, USA\\
    saikatd@cornell.edu
}
\and
\IEEEauthorblockN{Ziyang Li}
\IEEEauthorblockA{
    \textit{University of Pennsylvania}\\
    Philadelphia, USA\\
    liby99@seas.upenn.edu
}
\linebreakand
\IEEEauthorblockN{Alaia Solko-Breslin}
\IEEEauthorblockA{
    \textit{University of Pennsylvania}\\
    Philadelphia, USA\\
    alaia@seas.upenn.edu
}
\and
\IEEEauthorblockN{Rajeev Alur}
\IEEEauthorblockA{
    \textit{University of Pennsylvania}\\
    Philadelphia, USA\\
    alur@seas.upenn.edu
}
\and
\IEEEauthorblockN{Mayur Naik}
\IEEEauthorblockA{
    \textit{University of Pennsylvania}\\
    Philadelphia, USA\\
    mhnaik@seas.upenn.edu
}
}

\maketitle

\def\thefootnote{*}\footnotetext{Equal contribution}\def\thefootnote{\arabic{footnote}}

\begin{abstract}
Security vulnerabilities in modern software are prevalent and harmful. While
automated vulnerability detection techniques have made promising progress, their
scalability and applicability remain challenging. 
The remarkable performance of Large Language Models (LLMs), such as GPT-4 and \clama, on code-related tasks
has prompted recent works to explore if LLMs can be used to detect security vulnerabilities. 
In this paper, we perform a more comprehensive study by 
 examining a larger and more diverse set of datasets, languages, and LLMs,
and qualitatively evaluating detection performance across prompts and vulnerability classes.
Concretely,
we 
evaluate the effectiveness of 16 pre-trained LLMs on 5,000 code samples—1,000 randomly selected each from five diverse security datasets. These balanced datasets encompass synthetic and real-world projects in Java and C/C++ and cover 25 distinct vulnerability classes.

Our results show that LLMs across all scales and families 
show modest effectiveness in end-to-end reasoning about vulnerabilities, 
obtaining an average accuracy of 62.8\% and F1 score of 0.71 across all datasets.
LLMs are significantly better at detecting vulnerabilities that typically only need intra-procedural reasoning, such as OS Command Injection and NULL Pointer Dereference.
Moreover, LLMs report higher accuracies on these vulnerabilities than popular static analysis tools, such as CodeQL.

We find that advanced prompting strategies that involve step-by-step analysis significantly improve performance
of LLMs on real-world datasets in terms of F1 score (by up to 0.18 on average).
Interestingly, we observe that LLMs show promising abilities at performing parts of the analysis correctly, such as identifying vulnerability-related
specifications (e.g., sources and sinks) and leveraging natural language information to understand code behavior (e.g., to check if code is sanitized).
We believe our insights can motivate future work on LLM-augmented vulnerability detection systems.

\Comment{By designing a series of effective prompting strategies, we obtain the best
results on the synthetic datasets with GPT-4: F1 scores of 0.79 on \owasp, 0.86
on \julietj, and 0.89 on \julietc. Expectedly, the performance of LLMs drops on
the more challenging real-world datasets: \cvej and \cvec, with GPT-4 reporting
F1 scores of 0.48 and 0.62, respectively. }
\Comment{We show that LLMs can often outperform existing static analysis and
deep learning-based vulnerability detection tools, especially for certain
classes of vulnerabilities. Moreover, LLMs also often provide reliable
explanations, precisely identifying the vulnerable data flows in code.}

\end{abstract}


\section{Introduction}
Security vulnerabilities afflict software despite decades of advances in
programming languages, program analysis tools, and software engineering
practices. Even well-tested and critical software such as OpenSSL, a widely used
library for applications that provide secure communications, contains trivial
buffer overflow vulnerabilities, e.g.,~\cite{CVE-2022-3602}
and~\cite{CVE-2022-3786}. A recent study by Microsoft showed that more than 70\%
of vulnerabilities are still caused by well-understood memory safety
issues~\cite{msbugs}. 
This is alarming given the rapidly growing size and complexity
of modern software systems,
encompassing numerous programs, libraries, and modules
that interact with each other.
Hence, we need major technical advances to effectively
detect security vulnerabilities in such complex software.

Traditional techniques for automated vulnerability detection, such as
fuzzers~\cite{manes2018fuzzing}, and static analyzers such as
CodeQL~\cite{codeql} and Semgrep~\cite{semgrep} have made promising strides. For
example, in the last two years, researchers found over 300 security
vulnerabilities through custom CodeQL queries~\cite{codeqlqueries, zdibugs}.
However, these techniques face challenges in scalability and applicability.
Fuzzing requires manually crafted fuzz drivers
and does not scale to large critical programs with complex inputs, 
such as network servers, 
embedded firmware, 
and system services.
On the other hand, static analysis relies heavily on manual API specifications, and
skillfully crafted heuristics to balance precision and scalability. 
Until recently,
GitHub paid a bounty of over 7K USD for each CodeQL query that
found new critical security bugs~\cite{githubbounty}.

\begin{table*}[!htb]
    \centering
    \footnotesize
    \caption{Summary of our research questions and key findings}
    \setlength{\tabcolsep}{0.2em}
    \label{tab:summary}
    \begin{tabular}{|p{0.24\textwidth}|p{0.7\textwidth}H|} \toprule
        \textbf{Research Questions} & \textbf{Findings} & \textbf{Implications} \\ \midrule
    \textbf{RQ1:} How do different pre-trained LLMs perform in detecting security vulnerabilities across different languages and datasets? 
    (Section~\S\ref{sec:rq1}) &
    \positive LLMs across all sizes report a mean accuracy of about 62.8\%
    and a mean F1 score of 0.71 across all datasets. 
    
    \negative Average accuracy on real-world datasets is 10.5\% lower than that on synthetic datasets.
    
    \positive In stark contrast to other domains, smaller models such as \qwenfourteen and \qwenthirtytwo report
    higher accuracies on the real-world datasets than much larger models such as GPT-4.
    & 
    \textbf{For researchers:} There is room for improving vulnerability detection
    on real-world datasets. 
    While LLMs are useful for local reasoning and semantic understanding of code, future techniques should leverage these capabilities in conjunction with other tools (static analysis, etc.) to improve performance.

    \textbf{For developers:} LLMs should be used for vulnerability detection with caution by taking into account the heuristic discussed in other implications. 
    \\ \midrule
    
    \textbf{RQ2:} How do different prompting strategies affect the performance of LLMs? (Section~\ref{sec:rq2})  &
    \positive Using prompts that focus on detecting specific CWEs improves the performance of LLMs. 
    
    \positive The \dfstr prompt further improves results for larger LLMs with an increase of up to 0.18 in F1 score on real-world datasets.
    
    \positive We also observe that LLMs often infer the correct sources/sinks/sanitizers but fail in end-to-end reasoning. 
    &
    \textbf{For researchers:} Future approaches should combine LLMs with symbolic tools that can (a) handle the logical reasoning aspects of analysis and / or
    (b) provide more information about external calls / sources in large real-world projects.
    
    \textbf{For developers:} LLMs should be used to detect specific CWEs (as opposed to detecting \textit{any} vulnerability). 
    Moreover, \dfstr should be used for identifying sources/sinks and the downstream reasoning should be offloaded to off-the-shelf dataflow analysis tools.
    \\ \midrule
    
    \textbf{RQ3:} How does the performance of LLMs vary across different vulnerability classes? (Section~\ref{sec:rq3})  &
    \positive LLMs are better at detecting local vulnerabilities that require no global context across datasets (OS Command Injection, NULL Pointer Dereference, Out-of-bounds Read/Write, etc.).

    \negative LLMs struggle to detect vulnerabilities that require additional context or reasoning about complex data structures
    (Out-of-bounds Read/Write with C++ structs and pointers).

    \positive Certain LLMs are very good at detecting specific CWEs across datasets 
    (\llamaseventy on OS Command Injection, DeepSeekCoder-7B on NULL Pointer Dereference, etc.).
    & 
    \textbf{For researchers:} Future techniques should focus on language-specific and vulnerability-specific adaptation of LLMs. 
    
    \textbf{For developers:} LLMs (in isolation) should be avoided for detecting vulnerabilities that cannot be locally reasoned about.
    \\ \midrule

    \textbf{RQ4:} How do LLMs compare to state-of-the-art static analysis tools? (Section~\ref{sec:rq5}) &
   \negative LLMs report lower overall accuracies than CodeQL on all synthetic datasets.

   \positive LLMs report higher accuracies than CodeQL on certain vulnerability classes across datasets (Path Traversal, OS Command Injection, etc). 
   CodeQL reports higher accuracies on Integer Overflow across datasets.
   
    &
    \textbf{For researchers:} 
    Future techniques should leverage the ability of LLMs to perform well on certain vulnerability classes to improve static analysis-based vulnerability
    detection tools.
    \\ \midrule
    
    \textbf{RQ5:} How do LLMs compare to state-of-the-art deep-learning-based tools? (Section~\ref{sec:rq6}) &
    \positive Deep Learning(DL)-based tools such as DeepDFA~\cite{steenhoek2023dataflow} and LineVul~\cite{fu2022linevul} report accuracies similar to \qwenthirtytwo on \cvec 
    even after being trained on in-distribution samples whereas \qwenthirtytwo reports higher F1 scores.
    
    \positive DL-based tools report lower accuracies and F1 scores than LLMs when trained and evaluated on different datasets.
    
    \positive LLMs provide natural language explanations for their predictions while DL-based tools provide binary scores and line numbers that are often difficult to interpret.
    &
    \textbf{For researchers:} Since pre-trained LLMs match the performance of custom deep-learning-based tools without additional training, 
    new techniques should focus on improving vulnerability detection in real-world codebases by combining these models with existing code analysis systems.
    \\ \bottomrule
    \end{tabular}
\end{table*}

Large Language Models (LLMs), including pre-trained models such as GPT-4 and
CodeLlama, have made remarkable advances in code-related tasks in a relatively short period.
Such tasks include code completion~\cite{copilot}, automated program
repair~\cite{xia2023automated,joshi2023repair,xia2022less}, test
generation~\cite{lemieux2023codamosa,deng2023large}, code
evolution~\cite{zhang2023multilingual}, and fault
localization~\cite{yang2023large}. These results clearly show the promise of
LLMs, opening up a new direction for exploring advanced techniques. Hence, an
intriguing question is whether the state-of-the-art pre-trained LLMs can also be
used for detecting security vulnerabilities in code.

To develop LLM-based solutions, an important first step is to systematically evaluate
the ability of LLMs in detecting \emph{known} vulnerabilities.
This is especially important in light of the rapidly
evolving landscape of LLMs in three aspects: {\em scale}, {\em diversity}, and
{\em applicability}. First, scaling these models to ever larger numbers of
parameters has led to significant improvements in
their capabilities\cite{Wei2022EmergentAO}. 
For instance, GPT-4, which is presumably orders of
magnitude larger than its 175-billion predecessor GPT-3.5,
significantly outperforms GPT-3.5 on a wide range of code-understanding tasks
\cite{bubeck2023gpt4}. Second, the diversity of LLMs has grown rapidly and now
includes not only proprietary general-purpose ones such as GPT-4 but also open-source
code-specific LLMs 
such as CodeLlama~\cite{roziere2023code} and StarCoder~\cite{li2023starcoder}. 
Finally, the reasoning capabilities of LLMs (and hence their applicability) may vary
significantly across different prompting strategies and programming languages.
\Comment{Finally, the number of techniques for adapting LLMs to solve complex tasks is
also growing quickly, and includes prompting techniques to elicit reasoning
behaviors from LLMs~\cite{Wei2022ChainOT, Yao2022ReActSR}, as well as
fine-tuning to boost their performance on specific tasks.}
All these factors open up a large exploration space for applying LLMs to the
challenging task of vulnerability detection.

\begin{table}[t!hb]
    \centering
    \footnotesize
    \caption{Comparison with other studies that focus on vulnerability detection with LLMs. Superscript U indicates an unbalanced dataset. Static Analysis is abbreviated as SA and Deep Learning as DL.}
    \setlength{\tabcolsep}{0.1em}
    \label{tab:related-work-comparison}
    \begin{tabular}{l|c|c|c|c|c|c}
    \hline
        \textbf{Study Features} & \cite{zhou2024large} & \cite{gao2023far} & \cite{steenhoek2024comprehensive} & \cite{sp2024} & \cite{ding2024vulnerability} & \textbf{Our Work} \\ \toprule
        Languages & C/C++ & C/C++ & C/C++ & C,Py & C/C++ & C/C++,Java \\
        \#Samples & 368 & 347 & 100 & 96 & 25.9$\text{K}^{\text{U}}$ & 5000 \\
        \#CWEs & 25 & N/A & N/A & 8 & 140 & 25 \\ \midrule
        \#LLMs ($>$1B parameters) & 3 & 16 & 11 & 5 & 4 & 16 \\ \midrule
        Comparison of various LLMs& \cmark & \cmark & \cmark & \cmark & \cmark & \cmark \\
        Qualitative prompt analysis& \xmark & \xmark & \xmark & \cmark & \xmark & \cmark \\
        CWE-wise analysis& \xmark & \xmark & \xmark & \xmark & \xmark & \cmark \\
        Comparison with SA tools& \xmark & \cmark & \xmark & \xmark & \xmark & \cmark \\
        Comparison with DL tools& \cmark & \xmark & \xmark & \xmark & \cmark & \cmark \\
        \bottomrule
    \end{tabular}
\end{table}

\looseness=-1
\mypara{Our Work}
We study the vulnerability detection capabilities of
16 state-of-the-art LLMs across different scales and families,
including proprietary models such as Gemini and GPT-4,
and open-source models like CodeLlama and Qwen.
We evaluate these models on five popular security vulnerability datasets
across two languages, C/C++ and Java, and 25 vulnerability classes.

We first study how LLMs perform on the task of vulnerability detection
using three prompting strategies and how these strategies
qualitatively compare 
against each other. We also attempt to identify vulnerability classes
that most benefit from the use of LLMs and these prompting techniques.
Our simplest prompting
strategies include the \emph{\basic}, which simply asks an LLM to check for
any vulnerabilities in the given code and the \emph{\cwespecificstr}, which asks
the LLM to check for a specific class of vulnerabilities or CWEs (such as Buffer
Overflows). 
Inspired by the success of static analysis tools like CodeQL that use
dataflow analysis to detect vulnerabilities, 
we design a prompting strategy called \emph{\dfstr}.
This prompt asks the LLM to simulate a source-sink-sanitizer based dataflow
analysis on the target code snippet before predicting if it is vulnerable.

We next compare LLMs with existing vulnerability detection tools,
namely static analysis-based CodeQL and two deep learning-based 
techniques, DeepDFA~\cite{steenhoek2023dataflow} and LineVul~\cite{fu2022linevul}.
As discussed earlier, static vulnerability detection tools are limited 
by the need for concrete API specifications
and require compiling / building entire target projects before detection.
Pre-LLM deep learning-based approaches such as 
DeepDFA~\cite{steenhoek2023dataflow} and LineVul~\cite{fu2022linevul}
attempt to mitigate some of these limitations through fine-tuned neural
representations of code. 
On the other hand, 
LLMs do not require the compilation of complete projects as
they can be prompted to analyze partial code snippets.
Moreover, they already have an internal model of APIs through pre-training
and do not need to be trained on large datasets from scratch.
We analyze the benefits and shortcomings of LLMs over CodeQL, including
vulnerability classes where they outperform each other.
We also study how they compare against the deep learning based-approaches
in terms of generalization across datasets.

\mypara{Comparison with other studies}
There are other studies that also evaluate the effectiveness of LLMs on
the task of vulnerability detection~\cite{zhou2024large,gao2023far,steenhoek2024comprehensive,sp2024,ding2024vulnerability}. 
Table~\ref{tab:related-work-comparison} presents
a comparison of our work with these studies. 
We present our study as
the most comprehensive on this topic for the following reasons:

\begin{itemize}[leftmargin=*]
    \item \textbf{Size and diversity of the datasets:} 
    We curate a dataset of 5K samples, with equal number of vulnerable and non-vulnerable snippets.
    The only larger dataset is from \cite{ding2024vulnerability} but only 695 of their 25.9K samples are vulnerable. 
    Furthermore, our study is the first to include Java code. 
    \item \textbf{Comparison with non-LLM-based tools:} Our study is the first to compare LLMs with CodeQL
    and specialized Deep Learning-based tools. Moreover, we also find vulnerability classes where LLMs
    outperform CodeQL and vice versa which is useful for deployment.
    \item \textbf{Qualitative analysis of prompts and vulnerability classes:} While other studies quantitatively
    compare prompting strategies, we also attempt to qualitatively identify the benefits of various prompt elements.
    Furthermore, we identify partial capabilities offered by some of these prompts that can be leveraged in LLM-based detection tools.
    We also identify vulnerability classes where LLMs perform well.
\end{itemize}

\mypara{Contributions} 
Our research questions and key findings are summarized in Table~\ref{tab:summary}.
To summarize, we make the following contributions in this paper:
\begin{itemize}[leftmargin=*]
    \item{\textbf{Empirical Study:}}  We conduct the largest comprehensive study
    on how state-of-the-art LLMs perform in detecting security vulnerabilities 
    across 5000 samples from five
    datasets, two programming languages (C/C++ and Java) and covering 25 unique vulnerability classes.
    \Comment{, covering more than 5000 vulnerable methods and
    25 unique vulnerability classes (or CWEs).} 
    \item{\textbf{Comparison with other vulnerability detection tools}:} We contrast the 
    performance of LLMs against popular static analysis and deep-learning-based vulnerability
    detection tools. We also identify vulnerability classes where LLMs perform better/worse
    than some of these tools.
    \item{\textbf{Qualitative comparison of prompting strategies:}} 
    We quantitatively and qualitatively compare three prompting strategies, 
    including a novel prompt inspired by dataflow analysis-based vulnerability detection.
    \item{\textbf{Insights:}} We perform a rigorous manual analysis
    of LLMs' predictions and highlight vulnerability patterns that impact the
    performance of these models. 
\end{itemize}
\section{Approach}
\subsection{Datasets} \label{sec:datasets}
For our study, we select five diverse synthetic/real-world vulnerability datasets from two languages: C++ and Java. 
Table~\ref{tab:datasets} presents the details of each dataset, such as the dataset size, programming language, number of vulnerable and non-vulnerable samples, and the number of unique CWEs. 
The synthetic benchmarks, \emph{OWASP} and \emph{Juliet}, allow for easy comparison with CodeQL
and the real-world benchmarks, \emph{CVEFixes}, are useful for evaluating practical utility.
While many real-world datasets have been proposed in the literature, 
we selected \emph{CVEFixes} because it is the only dataset that
1) contains vulnerability \emph{metadata} such as CVE and CWE IDs, 
2) is \emph{two-sided}, i.e., contains both vulnerable and non-vulnerable code samples, and 
3) covers multiple languages such as Java and C/C++. 
Table~\ref{tab:rwdataset} shows a comparison of existing real-world vulnerability datasets. 
We briefly describe each of the selected datasets next:
\begin{table}[!htb]
    \centering
    \footnotesize
    \caption{Details of Selected Datasets}
    \label{tab:datasets}
    \vspace{-0.1in}
    \setlength{\tabcolsep}{0.5em}
    \begin{tabular}{llHrrr}\toprule
        \textbf{Dataset} & \textbf{Language} & \textbf{Type} & \textbf{Size} &
        \textbf{Vul/Non-Vul} &  \textbf{CWEs}   \\\midrule
        OWASP\cite{owasp} & Java  & Synthetic& 2740 & 1415/1325& 11\\
         SARD Juliet (C/C++)~\cite{julietcpp} & C/C++ & Synthetic &81,280 & 40,640/40,640& 118  \\
         SARD Juliet (Java)~\cite{julietjava} & Java & Synthetic & 35,940 & 17,970/17,970& 112 \\      
         CVEFixes~\cite{Bhandari2021CVEfixesAC}  & C/C++ &  Real &19,576 & 8223/11,347 & 131 \\
          CVEFixes~\cite{Bhandari2021CVEfixesAC}  & Java& Real & 3926 &1461/2465& 68  \\
         \bottomrule
    \end{tabular}
\end{table}

\subsubsection{OWASP (Synthetic)} The Open Web Application Security Project (OWASP)
benchmark~\cite{owasp} is a Java test suite designed to evaluate the
effectiveness of vulnerability detection tools. Each test represents a
synthetically designed code snippet containing a security vulnerability. 

\subsubsection{Juliet (Synthetic)} Juliet~\cite{boland2012juliet} is a widely-used vulnerability
dataset developed by NIST with thousands of synthetically
generated test cases from various known vulnerability patterns.

\subsubsection{CVEFixes (Real-World)} Bhandari et al.~\cite{Bhandari2021CVEfixesAC} curated a dataset, known as
CVEFixes, from 5365 Common Vulnerabilities and Exposures (CVE) records from the
National Vulnerability Database (NVD). From each CVE, they automatically
extracted the vulnerable and patched versions of each method in open-source
projects, along with extensive meta-data such as the corresponding CWEs, project
information, and commit data. 
These methods span multiple programming languages but we only consider the
C/C++ and Java methods in this work.

\begin{table}[!htb]
    \footnotesize
    \centering
    \caption{Comparison of Real-World Datasets}
    \setlength{\tabcolsep}{0.2em}
    \label{tab:rwdataset}
    \begin{tabular}{l|c|c|c|c} \toprule
        Dataset&Languages&CVE Metadata&Two-Sided&Multi-Lang\\ \midrule
        BigVul~\cite{bigvul}&C/C++&\cmark&\xmark&\xmark\\
        Reveal~\cite{chakraborty2021deep}&C/C++&\xmark&\xmark&\xmark\\          
        DiverseVul~\cite{Chen2023DiverseVulAN}&C/C++&\xmark&\cmark&\xmark\\
        DeepVD~\cite{wang2023deepvd}&C/C++&\xmark&\xmark&\xmark\\ \midrule
        CVEFixes~\cite{Bhandari2021CVEfixesAC}&C/C++, Java, ...&\cmark&\cmark&\cmark\\
        \bottomrule
    \end{tabular}
\end{table}

\subsection{Metrics}
\looseness=-1 To evaluate the effectiveness of each tool, we use the standard
metrics used for classification problems. In this work, a true positive
represents a case when a tool detects a true vulnerability. In contrast, a false
positive is when the tool detects a vulnerability that is not exploitable. True
and false negatives are defined analogously. We describe each metric in the
context of vulnerability detection.
\begin{itemize}[leftmargin=*]
    \item{\textbf{Accuracy:}} Accuracy measures how often the tool makes a
    correct prediction, i.e.,  whether a code snippet is vulnerable or not. It
    is computed as: $\frac{\textit{True Positives + True
    Negatives}}{\textit{\#Samples}}$.
    \item{\textbf{Precision:}} Precision represents what proportion of cases
    that a tool detects as a vulnerability is a correct detection. It is
    computed as:  $\frac{\textit{True Positives}}{\textit{True Positives + False
    Positives}}$.
    \item{\textbf{Recall:}} Recall represents what proportion of vulnerabilities
    the tool can detect. It is computed as: $\frac{\textit{True
    Positives}}{\textit{True Positives + False Negatives}}$.
    \item{\textbf{F1 score:}} The F1 score is a harmonic mean of precision and
    recall. It is computed as: $2*\frac{\textit{Precision *
    Recall}}{\textit{Precision + Recall}}$.
\end{itemize}

\subsection{Large Language Models}
We choose the most popular state-of-the-art pre-trained Large Language Models
(LLMs) for our evaluation. 
We choose three closed-source models (GPT-4, GPT-3.5 and \gemini)
and thirteen open-source models from the 
Codellama-x, Llama-3.1-x, Mistral-Codestral-x, DeepSeekCoder-x, Qwen2.5-x and Qwen2.5-Coder-x series.
We use the ``Instruct'' variants of the models wherever applicable
since they are fine-tuned to follow user instructions and hence can better adapt to specific reasoning tasks.
We access the GPT-x models and \gemini using the OpenAI and Google Gemini APIs respectively
and use the Hugging Face APIs~\cite{huggingface} to access the open-source models.
Table~\ref{tab:llmdetails} presents more details about the models.

\begin{table}[!htb]
\setlength{\tabcolsep}{0.2em}
\centering
\caption{Details of LLMs (increasing order of size)}
\label{tab:llmdetails}
\vspace{-0.1in}
{\footnotesize
\begin{tabular}{l|r|r|r} \toprule
     \textbf{Model} & \textbf{Model Version} & \textbf{Size} & \textbf{Context Size}  \\ \midrule 
     \qwenc & \texttt{qwen2.5-coder-1.5b} & 1.5B & 128K \\
     \qwencseven & \texttt{qwen2.5-coder-7b} & 7B & 128K \\
     \clamasvn & \texttt{Codellama-7b-instruct} & 7B & 16K \\
     \dscoderseven & \texttt{deepseekcoder-7b} & 7B & 4K \\
     \llamaeight & \texttt{llama-3.1-8b} & 8B & 128K \\
     \clamathr & \texttt{CodeLlama-13B-Instruct} & 13B & 16K \\
     \qwenfourteen & \texttt{qwen2.5-14b} & 14B & 128K \\
     \dscoderfifteen & \texttt{deepseekcoder-v2-15b} & 33B & 128K \\
     \codestral & \texttt{mistral-codestral-22b} & 22B & 32K \\
     \qwenthirtytwo & \texttt{qwen2.5-32b} & 32B & 128K \\
     \dscoderthirtythree & \texttt{deepseekcoder-33b} & 33B & 16K \\
     \clamathirtyfour & \texttt{CodeLlama-34B-Instruct} & 34B & 16K \\
     \llamaseventy & \texttt{llama-3.1-70b} & 70B & 128K \\
     \gemini & \texttt{gemini-1.5-flash} & N/A & 1M \\
     GPT-3.5 & \texttt{gpt-3.5-turbo-0613} & N/A & 4K\\ 
     GPT-4 & \texttt{gpt-4-0613} & N/A & 8K\\
     \bottomrule
\end{tabular}}
\end{table}

\subsection{Prompting Strategies for LLMs}
\label{sec:prompts}
We explore various prompting strategies that can assist LLMs in predicting if a
given code snippet is vulnerable. The LLMs discussed in this study support chat interactions with two major types of prompts: 
the \emph{system prompt} can be used to set the context for the entire conversation 
while \emph{user prompts} can be used to provide specific details throughout the chat session. 
We include a \emph{system prompt} at the start of each input to describe the task and
expected structure of the response. Since persona assignment has been shown to
improve the performance of GPT-4 on specialized tasks \cite{Salewski2023InContextIR}, we
add the line ``\emph{You are a security researcher, expert in detecting security
vulnerabilities}'' at the start of every system prompt to assign a persona of a
Security Researcher to the model. The system prompt for all experiments ends
with the statement ``\emph{Provide response only in the following format:}'' followed by
an expected structure of the response from the model. The system prompt
is followed by a \emph{user prompt} that varies across the various prompting
strategies. In all our experiments, we incorporate the target code snippet into
the user prompt without any changes.

We construct three different prompting strategies:

\subsubsection{\basic}
We design a very simple prompt (shown in Listing 4 in the Appendix)
to test if the model can take a target code snippet as
input and detect if it is vulnerable and determine the correct CWE as well.\Comment{We
refer to it as the \generic prompt in this study.} The prompt begins with the
message ``\emph{Is the following code snippet prone to any security vulnerability?}''
followed by the code snippet. 

\begin{table}[t!ht]
    \centering
    \setlength{\tabcolsep}{0.1em}
     \caption{Dataset Processing and Selection}
     \label{tab:processing}
    \footnotesize
    \setlength{\tabcolsep}{0.2em}
    \begin{tabular}{l|rrrrr|r}\toprule
    Steps& \owaspabv & Juliet & Juliet & CVEFixes & CVEFixes & Total\\    
    & & C/C++& Java& C/C++ & Java & \\\midrule
   Original &2740 & 128,198 & 56,162 & 19,576 & 3926 & 210,602\\
   Filtering & 2740 & 81,280 & 35,940 & 19,576 & 3926 & 144,002\\
   Top 25 CWE & 1478 & 11,766 & 8,506 & 12,062 & 1810&23,560 \\
   Random Selection & 1000 & 1000 & 1000 & 1000 & 1000 & 5000\\
         \bottomrule
    \end{tabular}
\end{table}
\begin{figure*}[tbh!]
\centering
\footnotesize
\includegraphics[width=\textwidth]{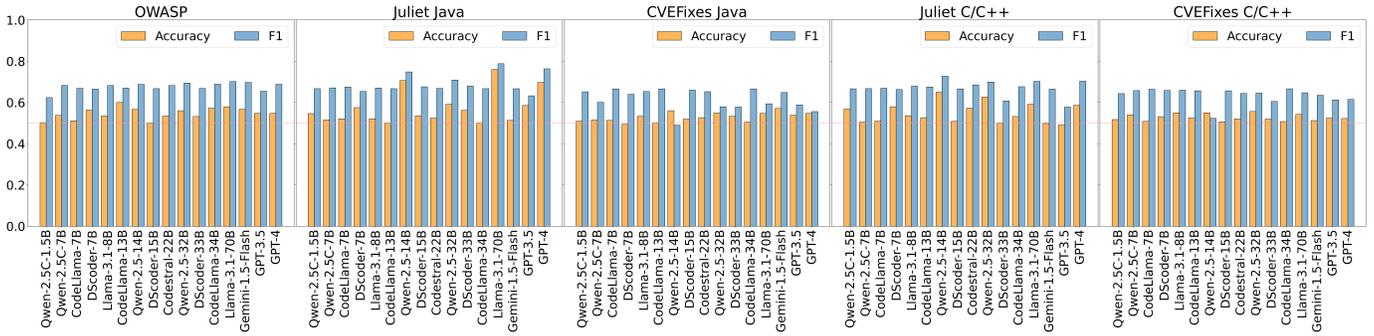}
\caption{Effectiveness of LLMs in Predicting Security Vulnerabilities in Java and C/C++ (highest accuracy and F1 scores per model per dataset, across all prompting strategies).}
\label{fig:best_metrics_per_model_per_dataset}
\vspace{-0.1in}
\end{figure*}
\subsubsection{\cwespecificstr}
The \cwespecificstr is presented in Listing 5 in the Appendix.
This prompt is similar to the \generic prompt except that it asks the model to
predict if the given code snippet is vulnerable to a specific target CWE.
Hence, the user prompt starts with ``\emph{Is the following code snippet prone to
$<$CWE$>$?}'' followed by the code snippet.\Comment{Same for other strings that are part of prompts.}
For instance, for CWE-476, the user prompt would start with ``\emph{Is the following code snippet
prone to CWE-476 (NULL Pointer Dereference)?}'' followed by the target code snippet.

\subsubsection{\dfstr}
Dataflow analysis 
is used by several static analysis tools to infer if there exists an unsanitized path from a source to a target node.
Further, prior literature has shown step-by-step instructions can 
often elicit better reasoning from LLMs~\cite{Wei2022ChainOT}.
Motivated by these observations, we designed the \df prompt (shown in
Listing 6 in Appendix) that prompts the model to simulate a
source-sink-sanitizer-based dataflow analysis on the target code snippet before
predicting if it is vulnerable.  Naturally, 
this prompt is costlier since it generates more tokens than the other prompts.
We provide the full prompts in Appendix B.

\subsubsection{Other prompting strategies}
We also tried other prompting strategies such as \texttt{Few-shot prompting} and
\texttt{Chain-of-thought prompting}. In the few-shot prompting setup, we include
two examples of the task (one with a vulnerability and one without) in the
\cwespecificstr before providing the target code snippet. 
With Chain-of-thought prompting, 
we prompt the model to generate a reasoning chain before the
final answer by adding a ``\emph{Let's think step-by-step}'' statement at the end of
the \cwespecificstr. 
Our initial experiments with GPT-4 prompted using these techniques
did not yield results better than the \dfstr on 100 random samples from two datasets.
Hence, we do not include these prompts in this study. 
We refer readers to Appendix 
C
for more details.

\subsection{Dataset Processing and Selection}
We preprocess each dataset before evaluation by 
removing or anonymizing information such as commits, benchmark IDs,
or vulnerability names that may provide obvious hints about the vulnerability. 
We also skip benchmarks that are spread across multiple files, due to limitations of prompt size.
We only consider samples corresponding to vulnerability types (CWEs)
listed in MITRE's Top 25 Most Dangerous Software Weaknesses~\cite{mitretop}.
We then filter code snippets with more than 2048 tokens due to prompt size limitations
and randomly sample 500 vulnerable and 500 non-vulnerable samples per dataset.
Table~\ref{tab:processing} presents the details of our selection stages.
We provide more details for each dataset in Appendix A. 

\subsection{Experimental Setup}
\mypara{Experiments with closed-source models} 
We use the OpenAI public API's \texttt{ChatCompletions} endpoint to
perform the experiments with GPT-3.5 and GPT-4. 
We use Google's Gemini API for the experiments with \gemini.

\mypara{Experiments with open-source models} 
We run all experiments with the open-source LLMs using the HuggingFace API 
on 
a cluster with A100, A6000, and RTX 2080 GPUs.

In all our experiments, we set the sampling temperature to 0 for
obtaining deterministic predictions, the maximum number of tokens to 1024, and
use the default values for all other parameters. 
We use the top-1 predictions for evaluation.

\section{Results}

\subsection{\textbf{RQ1:} Effectiveness of LLMs }
\label{sec:rq1}


\noindent 
We evaluate the performance of pre-trained LLMs on five open-source datasets
discussed in Section~\ref{sec:datasets}. 
Figure~\ref{fig:best_metrics_per_model_per_dataset} 
presents the best accuracy and F1 scores (across prompts) of the 16 models from Table~\ref{tab:llmdetails} on all datasets. 
The more detailed metrics for all prompts are presented in Appendix D.

\mypara{Finding 1.1: Modest Vulnerability Detection Performance Across LLMs} 
The best performing models and prompts per dataset report an accuracy of 62.8\% on average. 
The highest of these is reported by \llamaseventy (with \cwespecific) on the \julietj dataset (76\%).
The other best performing models per dataset are: 
\clamathr on \owasp (60\%), \gemini on \cvej (57\%), 
\qwenfourteen on \julietc (65\%) and \qwenthirtytwo on \cvec (56\%).
This confirms that no model individually performs the best across multiple datasets.
Moreover, 
the best accuracies on synthetic datasets are 10.5\% higher on average than those on the real-world datasets,
suggesting that these models might not be suitable for real-world vulnerability detection yet.

\mypara{Finding 1.2: Performance does not improve with scale}
On the real-world datasets, \qwenfourteen and \qwenthirtytwo report higher accuracies than the GPT-x models despite being much smaller.
We see many similar patterns across all models studied implying that model scale does not impact performance.
Within the same model class, the GPT-x models and the Llama-3.1-x models exhibit improvements in accuracy
as the size of the model increases on at least three datasets.
However, this is not observed in other model classes, i.e., the Qwen-2.5C-x, Qwen-2.5-x, Codellama-x and the DSCoder-x series.
This is in stark contrast to other domains where increasing model size leads to better performance.
Listing~\ref{lst:juliet1724} presents a representative example where 
\clamasvnabv correctly predicts that an integer overflow vulnerability (CWE-190) 
cannot occur in the given context while \clamathrabv does not.

\begin{figure}[!hbt]
\begin{minted}{java}
private void func() throws Throwable {
    int data;
    switch (5) {
    case 6:
        data = 0; break; // Hardcoded values in data
    default:
        data = 2; break; // Hardcoded values in data
    }
    switch (7) {
    case 7: // POTENTIAL FLAW: Integer Overflow 
        int result = (int)(data + 1); // But no overflow due to hardcoded values in data |\label{lst:line:intoverflow:add}|
        IO.writeLine("result: " + result);  break;
    default:
        IO.writeLine("fixed string"); break;
\end{minted}
\vspace{-0.3in}
\clamasvnprediction{
The code snippet does not
contain any obvious security vulnerabilities. \cmark
}
\vspace{-0.1in}
\clamathrprediction{
The integer overflow occurs when the value of `data'
exceeds the maximum value that can be stored in an integer, causing the value to
wrap around to a negative number... potentially allow an attacker to execute
arbitrary code. \xmark
}
\captionof{lstlisting}{\clamasvnabv correctly predicts this code is not vulnerable to Integer Overflow but \clamathrabv does not.}
\label{lst:juliet1724}
\end{figure}

\subsection{\textbf{RQ2:} Comparison of Prompting Strategies}
\label{sec:rq2}
\begin{figure}[b!]
\centering
\includegraphics[width=\linewidth]{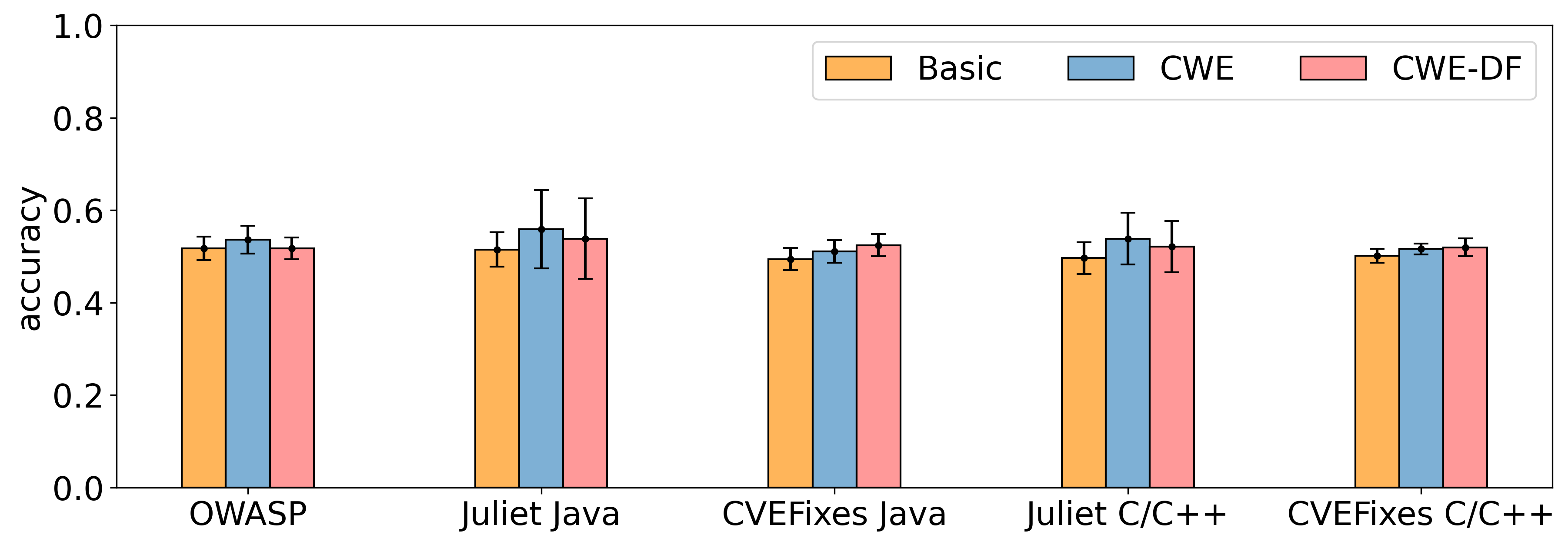}
\includegraphics[width=\linewidth]{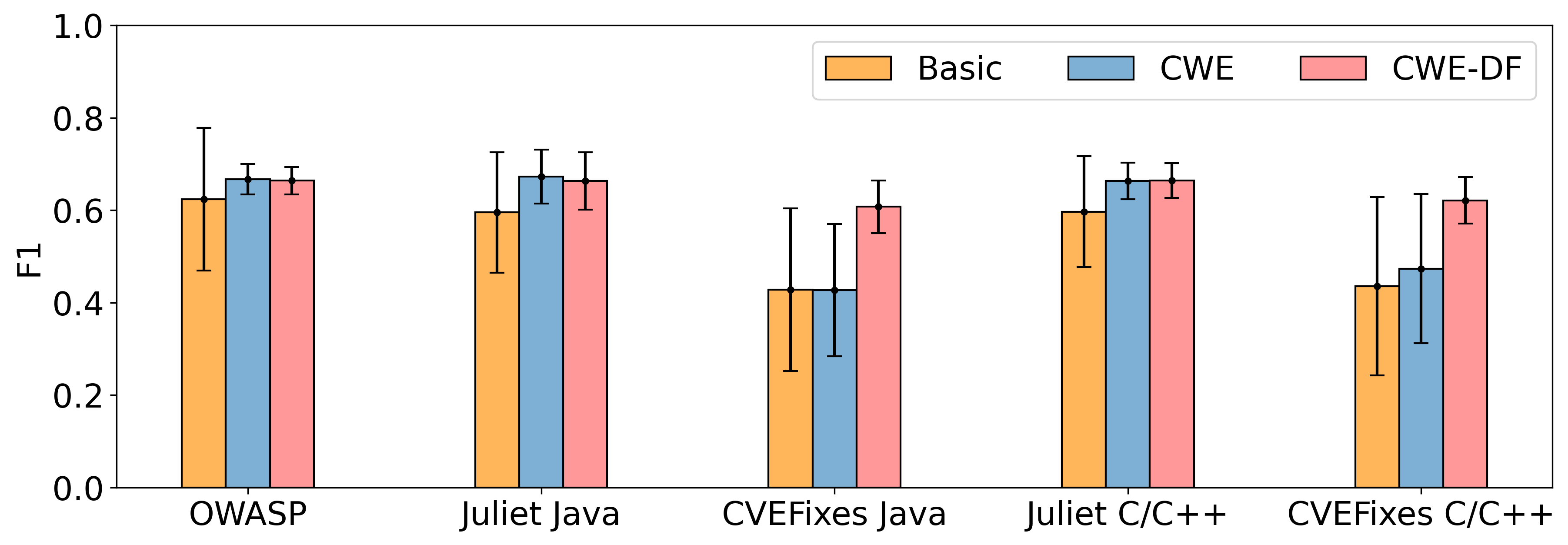}
\vspace{-0.1in}
\caption{Performance of different prompting strategies}
\label{fig:prompttrends}
\end{figure}

Figure~\ref{fig:prompttrends} presents the accuracy and F1 scores (averaged across all LLMs) 
of the three prompting strategies across all datasets.
Overall, the three prompts perform similarly in terms of accuracy on all datasets 
with \df reporting slightly higher accuracies on the real-world datasets.
Interestingly, the \df prompt reports significantly higher F1 scores on average than \cwespecific and \basic
on the real-world datasets (by upto 0.18 on \cvec and 0.14 on \cvej).
Furthermore, \df reports the lowest variance in accuracies and F1 scores across models,
as suggested by the standard deviation bars in Figure~\ref{fig:prompttrends}.
On the other hand, the {\basic} reports the highest variance in F1 scores.
We next highlight qualitative differences between prompts:

\mypara{Finding 2.1: Specifying the \cwespecific in the prompt reduces false alarms} 
Table~\ref{tab:basic_cwe} presents the percentage of samples where the \basic
predicts the correct CWE, averaged across datasets for 4 models.
We observe that the \basic detects incorrect CWEs in $>$ 60\% and $>$ 53\%
of all Java and C/C++ samples across models. 
We further manually inspected 10 vulnerable and 10 non-vulnerable samples from 
\julietj where GPT-4 with \basic is incorrect.
In only 4 or the 20 samples, the \basic predicts a plausible CWE.
However, these CWEs are unlikely due to the context. 
For example, it predicts that a value read from an input stream can be vulnerable if not validated (CWE-20) 
but this value is not used in a vulnerable context.
The \cwespecificstr (i.e., the \basic with CWE) improves 
or retains accuracy over the \basic on all 5 datasets.
GPT-4 with the \cwespecificstr not only correctly predicts all the 20 samples discussed above 
but also provides useful high-level explanations 
for why the snippet is vulnerable/not vulnerable in 18 / 20 samples. 
The 2 incorrect explanations are artifacts of faulty reasoning or hallucination: 
for example, an Integer Overflow due to addition to INT\_MAX in the function is 
incorrectly attributed to subtracting from INT\_MIN in the explanation.
Based on these observations, \emph{specifying the \cwespecific in the prompt can be helpful
in reducing incorrect predictions.}

\begin{table}[!htb]
    \centering
    \footnotesize
    \setlength{\tabcolsep}{0.4em}
    \caption{Correct CWEs detected with \basic (\%)}
    \label{tab:basic_cwe}
    \vspace{-0.1in}
    \begin{tabular}{l|ccccH}
\toprule
\textbf{Language (Avg.)} & \textbf{GPT-4} & \textbf{GPT-3.5} & \textbf{\clamathirtyfourabv} & \textbf{\clamathrabv} & \textbf{\clamasvnabv} \\ \midrule
 \textbf{Java} & 0.41  & 0.34 & 0.37 & 0.38 & 0.39  \\
 \textbf{C/C++} & 0.29 & 0.31 & 0.33 & 0.35 & 0.47 \\
\bottomrule
\end{tabular}
\end{table}

\mypara{Finding 2.2: Dataflow analysis identifies CWE-relevant textual cues and provides actionable explanations}
The \dfstr (\df) performs better than \cwespecificstr 
on the real-world datasets, \cvec and \cvej.
We inspect 10 vulnerable and 10 non-vulnerable samples from \cvej where 
GPT-3.5 
correctly predicts only with \df.
\emph{We find that the \df prompt leverages textual cues for sanitization 
(e.g., csrftokenhandler() suggests protection from CSRF) in 16/20 samples
that are missed by \cwespecificstr.} 

Further, \emph{\df responses are more useful in localizing the vulnerability  
as it correctly predicts the correct sources and sinks in 18 / 20 samples, 
sanitizers in 16 / 20 samples, and unsanitized flows in all samples}. 
Listing~\ref{fig:motivating-example} presents a response from GPT-4 using \df prompt 
that correctly identifies the unsanitized flows between sources and sinks. 
We present more \df examples in Appendix F. 

\mypara{Finding 2.3: \df identifies the correct sources and sinks even when the final prediction is incorrect}
We also inspect 10 vulnerable and 10 non-vulnerable samples 
from \julietc where \df's predictions are incorrect. 
Surprisingly, the sources, sinks and sanitizers are correctly identified in
17 / 20 samples but the unsanitized flows are incorrect in 12 / 20 samples.
Hence, the final predictions are incorrect only due to erroneous reasoning about the snippet / 
false assumptions about the CWE. 
Listing~\ref{fig:julietcpp43583} presents an example where the vulnerability is not detected
but the sources and sinks are correctly identified.
This suggests that \emph{the \df prompt can be used to accurately identify
sources/sinks/sanitizers and other dataflow analysis techniques can be used to reason
about unsanitized flows to predict the vulnerability.}

\begin{figure}[t!]
\begin{minted}{c}
// Code snippet prone to CWE-78 (OS Command Injection)
// Unverified data from socket is used as an argument to _execvp
void func(){
    char dataBuffer[100] = "dir "; 
    char *data = dataBuffer; size_t dataLen = strlen(data);
    int recvResult; SOCKET connectSocket = INVALID_SOCKET;
    do {// setup socket and receive user input 
        conn = socket(AF_INET, SOCK_STREAM, IPPROTO_TCP);   |\label{lst:line:eg:socket}|
        recvResult = |\colorbox{yellow}{recv}|(conn, (char *)(data + dataLen), sizeof(char) * (100 - dataLen - 1), 0)}; |\label{lst:line:eg:recv}|
        // missing validation of recvResult
        data[dataLen + recvResult / sizeof(char)] = '\0'; |\label{lst:line:eg:data}|
    } while (0)
    char *args[] = {"%WINDIR%\\system32\\cmd.exe", "/c", |\colorbox{yellow}{data}|, NULL}; |\label{lst:line:eg:args}|
    |\colorbox{pink}{\_execvp}|("cmd.exe", args); |\label{lst:line:eg:exec}|
}
\end{minted}
\vspace{-0.2in}
\prediction{The data received from the socket connection is directly used in the \texttt{\_execvp} function without proper sanitization. This could allow an attacker to inject malicious commands. \cmark}
\vspace{-0.2in}
\captionof{lstlisting}{GPT-4 (\df) detects that this snippet is prone to OS Command Injection
due to unsanitized paths from a \colorbox{yellow}{source} to \colorbox{pink}{sink}.
CodeQL does not detect this vulnerability.}
\label{fig:motivating-example}
\vspace{-0.1in}
\end{figure}

\begin{figure*}[!hbt]
\centering
\footnotesize

\begin{subfigure}[b]{0.64\textwidth}   
    \includegraphics[width=\textwidth]{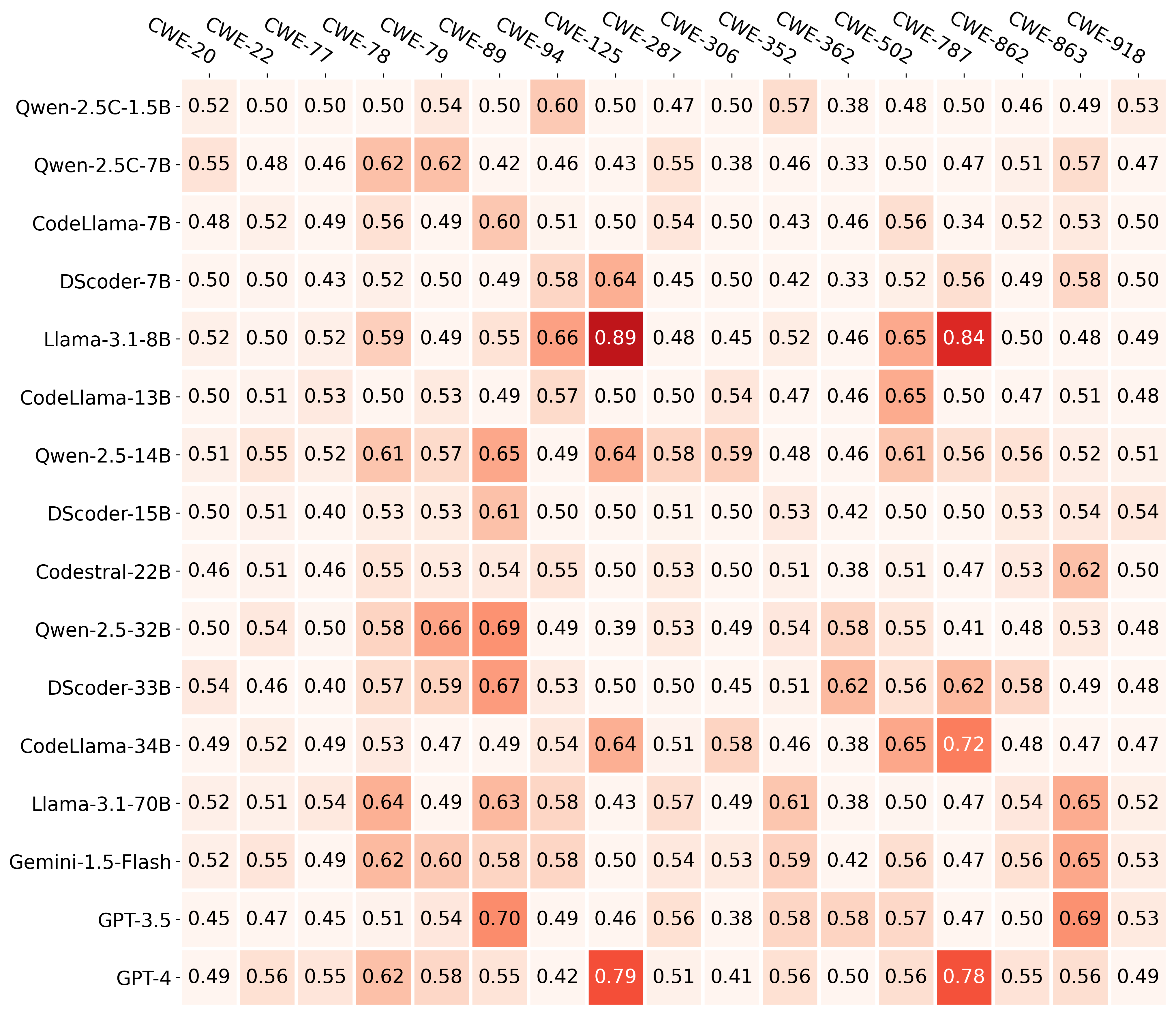}
    \caption{\cvej}
    \label{fig:gpt4_df_owasp_java_cwe_distribution}
\end{subfigure}
     \hfill
\begin{subfigure}[b]{0.345\textwidth}
    \includegraphics[width=\textwidth]{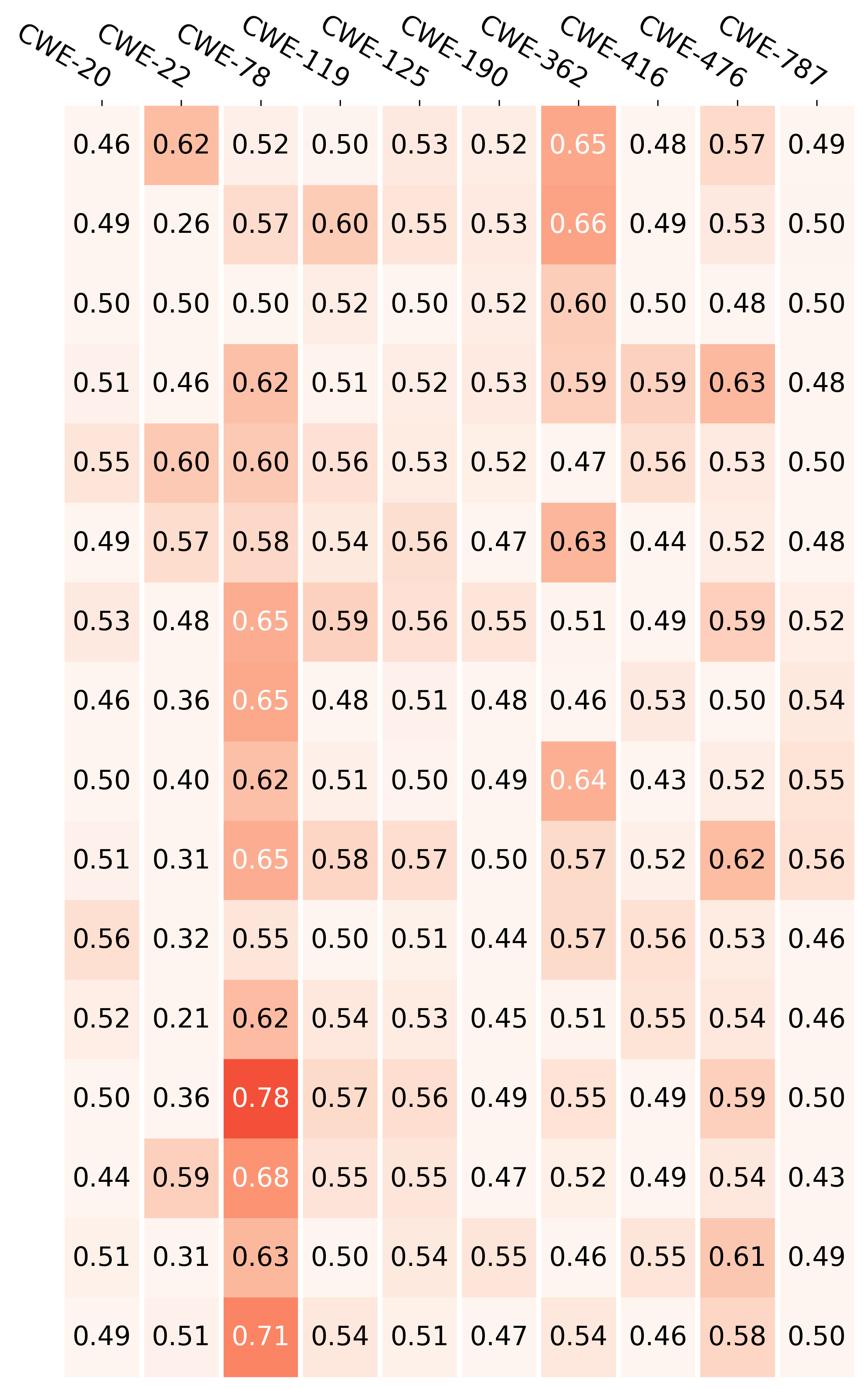}
    \caption{\cvec}
    \label{fig:gpt4_juliet_cpp_cwe_distribution}
\end{subfigure}
\caption{Accuracy Across CWEs on real-world datasets.}
\label{fig:java_cwe_distribution}
\end{figure*}

\subsection{\textbf{RQ3:} Performance of LLMs across CWEs}
\label{sec:rq3}


We next evaluate how the LLMs perform on different classes of security vulnerabilities (CWEs).
Because the CWE-wise distribution of vulnerable and non-vulnerable samples can be imbalanced, we compute balanced accuracy for each CWE (for ease of presentation, we refer to it as accuracy henceforth in this section). 
For each dataset and model, we consider the best-performing prompt for the analysis and only report CWEs with at least 10 samples. 
Figure~\ref{fig:java_cwe_distribution} presents the
CWE-wise distribution of accuracies on the real-world datasets,
\cvej and \cvec.
Figure~\ref{fig:cpp_cwe_distribution} reports the accuracies on the
synthetic datasets, \owasp, \julietj and \julietc.

\mypara{Finding 3.1: LLMs perform well on vulnerabilities that do not require additional context}
These CWEs include
\emph{OS Command Injection (CWE-78),
Out-of-bounds Read / Write (CWE-125, CWE-787), 
Null Pointer Dereference (CWE-476),
Cross-site Scripting (CWE-79),
SQL Injection (CWE-89)
and 
Incorrect Authorization (CWE-863).}
The higher
performance on these vulnerabilities can be attributed to the fact that these
are fairly self-contained and little additional context is needed to detect
them.
For example, 4/16 LLMs report accuracies $>$ 60\% on Incorrect Authorization (CWE-863) on \cvej 
it is easier to
validate if an implemented authorization check is correct or not.
On the other hand, no LLMs report high accuracies on Missing Authorization (CWE-862) since it's not known if an input parameter has been authorized outside the context of the target method and more context is hence needed to detect this vulnerability class.
The following summarizes how LLMs perform on these CWEs: 
\begin{itemize}[leftmargin=*]
    \item OS Command Injection (CWE-78) sees the highest performance across models and datasets with $>$60\% accuracies reported by 5/16 LLMs on \cvej and 11/16 LLMs on \cvec. \llamaseventy reports the best performance on CWE-78 with accuracies 64\% and 78\% on \cvej and \cvec respectively.
    \item \llamaeight and GPT-4 perform extremely well on
    Out-of-bounds Write (CWE-787) with accuracies of 89\% and 79\% on \cvej and \cvec respectively and on Out-of-bounds Read (CWE-125)
    with accuracies of 84\% and 78\% respectively. Moreover,
    accuracies over 60\% are reported on \cvej by 5/16 LLMs
    for CWE-125 and 4/16 LLMs for CWE-787.
    \item NULL Pointer Dereference (CWE-476) sees accuracies
    higher than 60\% by 3/16 LLMs on \cvec, 7/16 on \julietc
    and 11/16 on \owasp. Interestingly, \dscoderseven performs the best on all three datasets with accuracies of 63\%, 88\% and 83\% respectively.
    \item GPT-3.5 reports the highest accuracy of 70\% on SQL Injection (CWE-89) on \cvej and 7/16 LLMs report accuracies over 60\%. Surprisingly, all LLMs report accuracies $<$60\% on the same CWE on synthetic datasets (\owasp and \julietj).
    \item \qwenthirtytwo reports high accuracies of 67\% and 66\% 
    on Cross-Site Scripting (CWE-79) on \owasp and \cvej respectively.
    Accuracies over 60\% are reported by 3/16 LLMs on \cvej and 11/16 LLMs on \owasp.
\end{itemize}

\mypara{Finding 3.2: Poor performance on real-world C/C++ is due to missing global context}
We see that the performance of all LLMs on vulnerabilities in \cvec is
worse than that on the same CWEs in \cvej and \julietc.
While 
GPT-4 and \llamaeight perform extremely
well on the Out-of-bounds Read / Write vulnerabilities in \cvej
as discussed above,
they report accuracies $<$ 53\% for these CWEs on the \cvec dataset. 
In fact, no LLMs report accuracies $>$ 60\% for these CWEs on \cvec.
We attribute this disparity to the nature of these vulnerabilities 
in the two languages: \emph{Out-of-bounds Reads / Writes in \cvec require
reasoning about pointers and structs which requires more context about
the structs and their members.} In \cvej, on the other hand, 
these vulnerabilities arise primarily due to illegal array indexing.
This issue does not emerge in \julietc because all the information about the
pointers is presented in the snippet. We present examples in Appendix G.

\mypara{Finding 3.3: Some LLMs are better at detecting certain CWEs}
Concretely, the following LLMs report the best accuracies on certain CWEs across datasets:
\begin{itemize}[leftmargin=*]
    \item \llamaseventy on OS Command Injection (CWE-78)
    \item \dscoderseven on NULL Pointer Dereference (CWE-476)
    \item \qwenthirtytwo on Cross-Site Scripting (CWE-79)
    \item \llamaeight and GPT-4 on Java Out-of-bounds Read/Write (CWE-125/787)
\end{itemize}

\subsection{\textbf{RQ4:} LLMs vs Static Analysis Tools}
\label{sec:rq5}

\mypara{Experiment setup} 
We next explore how the LLMs compare against CodeQL.
Since CodeQL requires building projects before analysis and the real-world datasets contain large projects, 
we limit our focus to the three synthetic datasets, namely \owasp, \julietj and \julietc.
We use the official CodeQL queries designed for the top 25 CWEs.
Table~\ref{tab:codeql} presents results from CodeQL and
the best performing LLM on the three datasets:
\clamathr on \owasp, \llamaseventy on \julietj and \qwenfourteen on \julietc.

\mypara{Finding 4.1: CodeQL performs better than the LLMs in terms of accuracy on all three datasets}
CodeQL reports 0.07
and 0.15 higher F1 than \clamathr on \owasp and \llamaseventy on \julietj respectively.
However, \qwenfourteen reports a 0.11 higher F1 on \julietc.

\mypara{Finding 4.2: LLMs perform better than CodeQL on certain CWEs}
A CWE-wise comparison of LLMs with CodeQL in Figure~\ref{fig:cpp_cwe_distribution} reveals that
\emph{LLMs report higher accuracies on CWE-22 (Path Traversal), CWE-78 (OS Command Injection),
CWE-476 (NULL Pointer Dereference), 416 (Use After Free) on at least 2 / 3 datasets while
 CodeQL performs better on CWE-190 (Integer Overflow) on 2 datasets.}
 Concretely, 
\begin{itemize}[leftmargin=*]
    \item CodeQL performs better than the LLMs on CWE-190 (Integer Overflow) with
    11\% and 21\% higher accuracies than \llamaseventy on \julietj and \julietc respectively.
    \item On the other hand,
    \dscoderseven performs better than CodeQL on CWE-476 (NULL Pointer Dereference) 
    with 12\% higher accuracy on \julietj and only 1\% lower accuracy on \julietc.
    Moreover, 6 / 16 LLMs report accuracies higher than CodeQL on CWE-476 from \julietj.
    \item While CodeQL reports an extremely high accuracy of 95\% on CWE-78 (OS Command Injection) with \julietj ,
    it is outperformed by \clamathr by 1\% and \codestral by 7\% on \owasp and \julietc respectively.
    Interestingly, 7 / 16 LLMs report higher accuracies (by upto 10\%) at detecting CWE-22 (Path Traversal) on \owasp.
    \item Similarly, 3 LLMs perform better on CWE-416 (Use After Free) from \julietc 
    with \dscoderseven reporting a  whopping 24\% higher accuracy than CodeQL.
\end{itemize}

\begin{figure}[!thb]
\centering
\footnotesize
\definecolor{cornflowerblue86138221}{RGB}{106, 158, 231}
\definecolor{darkgray176}{RGB}{176,176,176}
\definecolor{peru21013090}{RGB}{225, 170, 116}

\begin{subfigure}[b]{0.415\linewidth}
    \includegraphics[width=\textwidth]{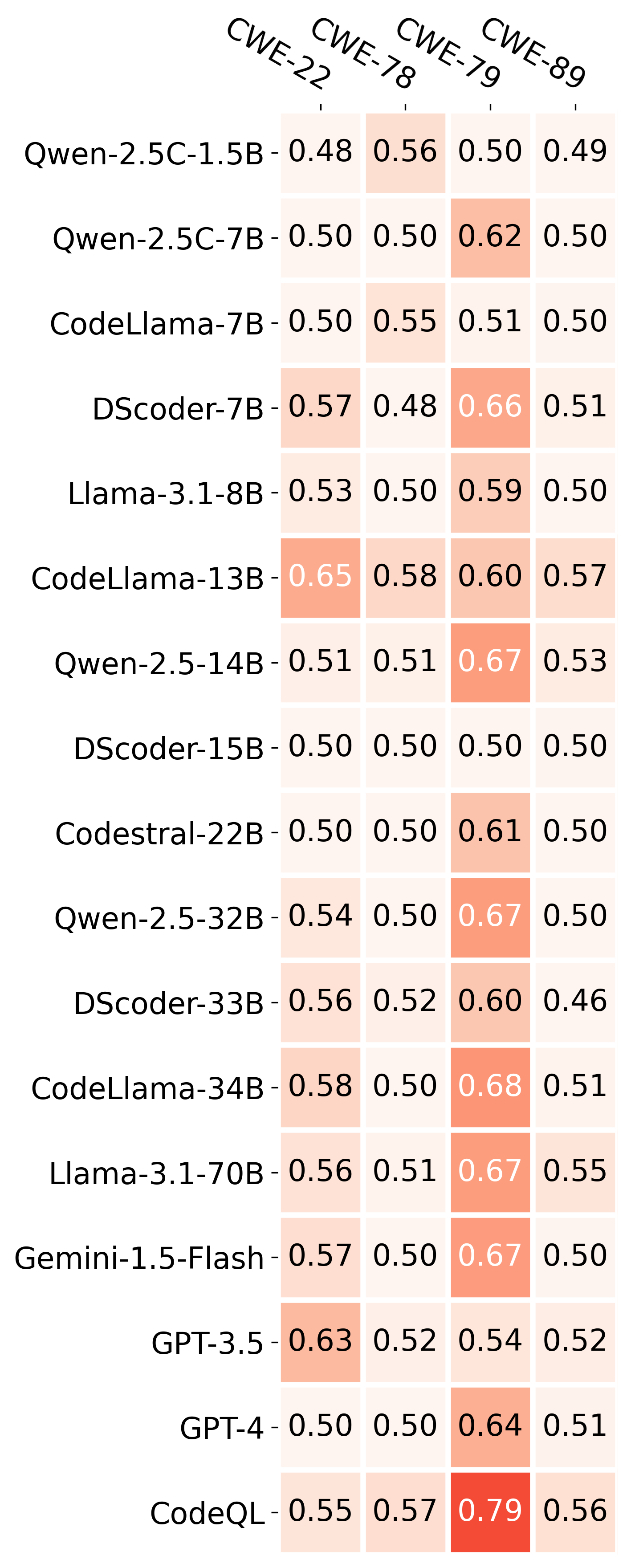}
    \caption{\owasp}
    \label{fig:owasp_cwe_distribution}
\end{subfigure}
\begin{subfigure}[b]{0.28\linewidth}
    \includegraphics[width=\textwidth]{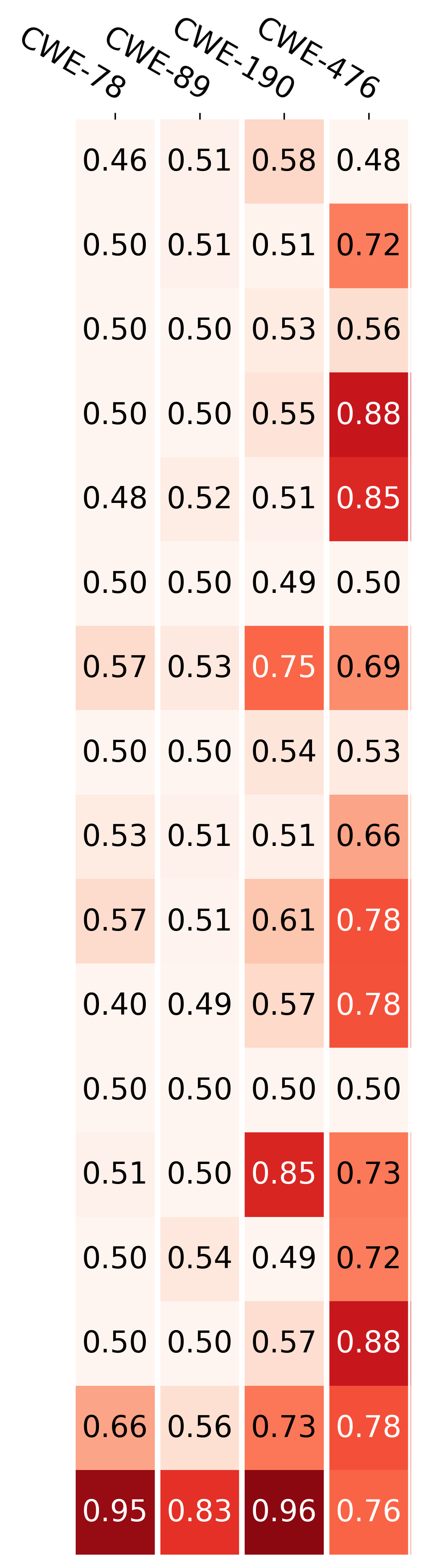}
    \caption{\julietj}
    \label{fig:juliet_java_distribution}
\end{subfigure}
\begin{subfigure}[b]{0.28\linewidth}
    \includegraphics[width=\textwidth]{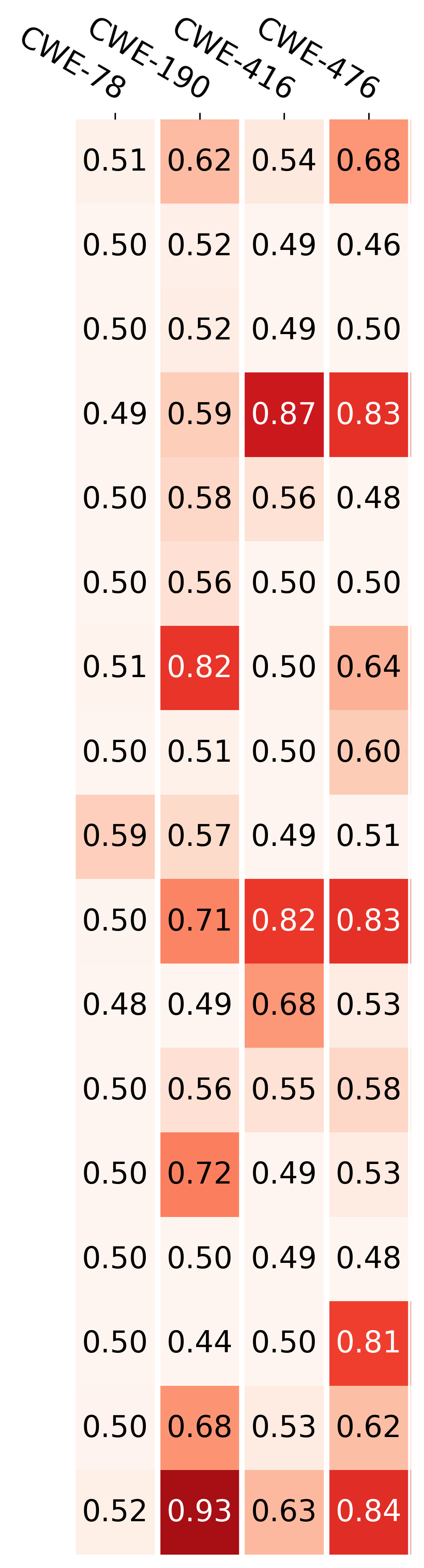}
    \caption{\julietc}
    \label{fig:juliet_cpp_distribution}
\end{subfigure}
\caption{Accuracy Across CWEs on synthetic datasets.}
\label{fig:cpp_cwe_distribution}
\vspace{-0.1in}
\end{figure}

\mypara{Finding 4.3: CodeQL's worse performance on some CWEs can be attributed to the very specific manually-written queries 
which may not cover all possible scenarios of the vulnerability}
For example, CodeQL only detects CWE-78 (OS Command Injection) in C/C++ snippets 
when there exist system commands that take a string of arguments (\texttt{execl}).
This query cannot handle commands that take a list of arguments (eg., \texttt{\_execvp}).
Listing~\ref{fig:motivating-example} provides an example of this scenario where
CodeQL does not detect that the snippet is prone to OS Command Injection
but GPT-4 (\df) correctly identifies \texttt{\_execvp} as a vulnerable sink.
Listing~\ref{fig:julietcpp43583} presents an example where 
CodeQL correctly predicts that the target snippet is vulnerable to Integer Overflow
while GPT-4 with \df does not. 
However,
the model correctly identifies the sources, sinks and even unsanitized dataflows in this case
but fails to faithfully reason over them when predicting the vulnerability.
These examples support the observation from Section~\ref{sec:rq2} that
\emph{LLMs can be used to identify sources and sinks relevant to the target vulnerability
(which can be missed by manually written queries) while
static analysis tools like CodeQL can be used to reason over them}.
More examples are presented in the Appendix E.

\begin{table}[!htb]
    \centering
    \footnotesize
    \caption{CodeQL vs. best performing LLM on synthetic datasets (\clamathr on \owasp, \llamaseventy on \julietj, and \qwenfourteen on \julietc)}
    \setlength{\tabcolsep}{0.5em}
    \label{tab:codeql}
    \vspace{-0.1in}
    \begin{tabular}{l|rrrr|rrrr}
\toprule
\textbf{Dataset/Model} & \multicolumn{4}{c|}{\textbf{CodeQL}} & \multicolumn{4}{c}{\textbf{Best performing LLM}} \\ \midrule
              &    A &    P &    R &   F1 &    A &    P &    R &   F1 \\ \midrule 
 \textbf{OWASP}        & 0.63 & 0.58 & 0.96 & 0.72 & 0.60 & 0.58 & 0.74 & 0.65 \\
 \textbf{Juliet Java}  & 0.94 & 0.92 & 0.95 & 0.94 & 0.76 & 0.71 & 0.89 & 0.79 \\
 \textbf{Juliet C/C++} & 0.72 & 0.99 & 0.44 & 0.61 & 0.65 & 0.60 & 0.89 & 0.72 \\
\bottomrule
\end{tabular}
\vspace{-0.1in}
\end{table}

\begin{figure}[!htb]
\begin{minted}{c}
// Code snippet prone to CWE-190 (Integer Overflow)
// data is incremented without verifying if data < INT_MAX

static const int STATIC_CONST_FIVE = 5;
void f7909()
{
    int64_t data;
    data = 0LL;
    if(STATIC_CONST_FIVE==5)
    {
        data = (int64_t)|\colorbox{yellow}{RAND64()}|;
    }
    if(STATIC_CONST_FIVE==5)
    {
        {
            ++data;
            int64_t result = data;
            |\colorbox{pink}{printLongLongLine}|(result);
        }
    }
}
\end{minted}
\vspace{-0.1in}
\prediction{\\
\textbf{Unsanitized Data Flows} \cmark\\
The data from \texttt{RAND64()} is directly used in \texttt{printLongLongLine()} without any sanitization or checks for integer overflow.)\\

\textbf{Vulnerability analysis verdict} \xmark\\
vulnerability: NO (CWE-190) \\

\textbf{Explanation} \xmark\\
The code does not perform any arithmetic operations that could lead to an integer overflow. The only operation performed on the \texttt{data} variable is incrementing it by one \texttt{(++data)}, which is unlikely to cause an integer overflow given that data is a 64-bit integer.
}
\vspace{-0.1in}
\captionof{lstlisting}{CodeQL correctly predicts Integer Overflow while GPT-4 doesn't. 
GPT-4, however, correctly identifies the \colorbox{yellow}{sources} and \colorbox{pink}{sinks} and the unsanitized paths.}
\label{fig:julietcpp43583}
\vspace{-0.1in}
\end{figure}

\subsection{\textbf{RQ5:} LLMs vs Deep Learning-Based Tools} 
\label{sec:rq6}
We compare LLMs against two prior deep learning-based approaches: 1) DeepDFA~\cite{steenhoek2023dataflow}, which trains Graph Neural Networks using embeddings of control flow graphs and associated data flow facts, and 2) LineVul~\cite{fu2022linevul}, which is a transformed-based model trained using token-based representation of code. 

\mypara{Experiment setup} 
Our main aim in this experiment is to check the generalizability of these techniques on real-world datasets
beyond the datasets they are trained on.
We use \cvec with the 1000 samples from our main evaluation~\S\ref{sec:rq1} as the real-world test set.
We train on three different datasets:\julietc, \cvec (excluding samples in the test set) and BigVul~\cite{bigvul}(the original C/C++ dataset that these models were trained on). 
We use an 80/20 train-validation split while training on these datasets.
We use the DeepDFA and LineVul versions from DeepDFA's latest artifact~\cite{deepdfaurl}. 
Table~\ref{tab:dltools} presents the results, averaged across three runs.

\mypara{Finding 5.1: DL-based approaches have limited effectiveness on real-world datasets} 
DeepDFA and LineVul trained on \cvec training set report accuracies of 51\% and 59\% on \cvec test set respectively while
the best performing LLM on this dataset, \qwenthirtytwo, reports an accuracy of 56\%.
This is in stark contrast to the performance of these techniques on BigVul where they report accuracies higher than 98\%.
Moreover, \qwenthirtytwo reports an F1 score of 0.65 which is 0.42 and 0.04 higher than DeepDFA and LineVul respectively.

\mypara{Finding 5.2: DL-based approaches do not generalize across datasets}
When trained on \julietc or BigVul, both DeepDFA and LineVul report accuracies and F1 scores lower than \qwenthirtytwo
by upto 6\% and 0.63 respectively.

\mypara{Finding 5.3: LLMs are preferable over DL-based approaches due to low inference overheads and natural language explanations}
DeepDFA involves significant inference overhead, due to the CFG extraction and dataflow analysis steps. LLMs, however, can use the textual representation of code and can operate on incomplete/partial programs. The use of data-flow and control-flow information in DeepDFA is evidently useful. We made similar observations with LLMs when using the \df prompt. 
On the other hand, LineVul, like LLMs, can leverage natural language information but has a generalization problem.
Finally, both DeepDFA and LineVul provide binary labels and line numbers that are difficult to interpret. LLMs can additionally provide explanations, which are useful for further debugging (as shown in prior sections). 

\begin{table}[t!hb]
    \centering
    \footnotesize
     \caption{\qwenthirtytwo vs DeepDFA vs LineVul on \cvec}
    \label{tab:dltools}
    \vspace{-0.1in}
     \setlength{\tabcolsep}{0.3em}
\begin{tabular}{lll|rrrr}
\toprule
 \textbf{Model} & \textbf{Train/Prompt}    & \textbf{Test}     & \textbf{A}           & \textbf{P}          & \textbf{R}          & \textbf{F1}            \\
\midrule 
 DeepDFA & \bigvul   & \bigvul   & 0.98 & 0.53 & 0.92  & 0.67   \\ 
  LineVul&\bigvul&\bigvul&0.99&0.96&0.88&0.92\\ 
 \midrule \midrule
 \qwenthirtytwo&\df&\cvec&0.56&0.54&0.81&0.65\\  
 
 \midrule
 DeepDFA &\cvec & \cvec & 0.51   & 0.55     & 0.17   & 0.23   \\
     DeepDFA & \julietc   & \cvec & 0.53   & 0.53   & 0.65   & 0.58   \\ 
  DeepDFA & \bigvul   & \cvec & 0.52  & 0.52 & 0.76 & 0.62 \\
 \midrule
LineVul&\cvec&\cvec&  0.59&0.58&0.65&0.61\\ 
LineVul&\julietc&\cvec& 0.50&0.50&0.91&0.64\\
LineVul&\bigvul&\cvec&0.50&0.63&0.01&0.02\\
\bottomrule
\end{tabular}
\end{table}

 \section{Related Work}

\looseness=-1
\mypara{Static analysis tools for vulnerability detection} 
Tools such
as FlawFinder~\cite{flawfinder} and CppCheck~\cite{cppcheck}  use syntactic and
simple semantic analysis techniques to find vulnerabilities in C++ code. More
advanced tools like CodeQL \cite{codeql}, Infer~\cite{fbinfer}, and
CodeChecker~\cite{codechecker} employ semantic analysis techniques and can
detect vulnerabilities in multiple languages. Static analysis tools rely on
manually crafted rules and precise specifications of code behavior, which is
difficult to obtain automatically. In contrast, while LLMs cannot always reliably
perform end-to-end reasoning over code, 
we find that they are capable of automatically identifying these specifications
which can be leveraged to improve static analysis-based detection tools.

\looseness=-1
\mypara{Deep Learning-based vulnerability detection} 
Several works have focused on using Deep Learning techniques for vulnerability
detection. Earlier works such as Devign~\cite{Zhou2019DevignEV},
Reveal~\cite{Chakraborty2020DeepLB},  LineVD~\cite{Hin2022LineVDSV} and
IVDetect~\cite{Li2021VulnerabilityDW} leveraged  Graph Neural Networks (GNNs)
for modeling dataflow graphs, control flow graphs, abstract syntax trees and
program dependency graphs. Other works explored alternate model
architectures: VulDeePecker~\cite{Li2020VulDeeLocatorAD} and
SySeVR~\cite{Li2018SySeVRAF} used LSTM-based models on slices and data
dependencies while Draper used Convolutional Neural Networks. Recent works
demonstrate that Transformer-based models fine-tuned on the task of
vulnerability detection can outperform specialized techniques (CodeBERT,
LineVul~\cite{fu2022linevul}, UnixCoder). DeepDFA~\cite{steenhoek2023dataflow}
and ContraFlow~\cite{Cheng2022PathsensitiveCE} learn specialized embeddings that
can further improve the performance of Transformer-based vulnerability detection
tools. These techniques, however, provide binary labels for
vulnerability detection and do not provide natural language explanations.

\mypara{LLMs for automated software engineering} Recent approaches have
demonstrated the effectiveness of LLMs
on software engineering tasks
such as automated program
repair~\cite{xia2023automated,joshi2023repair,xia2022less}, test
generation~\cite{lemieux2023codamosa,deng2023large}, code
evolution~\cite{zhang2023multilingual}, and fault
localization~\cite{yang2023large}. 
However, unlike these approaches, we find that 
scaling LLMs to larger sizes does not improve vulnerability detection abilities.
\cite{Thapa2022TransformerBasedLM} explore whether Language Models fine-tuned
on multi-class classification can perform well where the classes correspond to
groups of similar types of vulnerabilities. In contrast,  we study whether LLMs can perform a much granular CWE-level
classification through prompting.
Recent work combining LLMs with static analysis to detect Use Before Initialization (UBI) bugs in the Linux Kernel~\cite{li2024enhancing}
supports our claims in Section~\ref{sec:rq5} (but focuses on a specific class of bugs).
There are other concurrent studies evaluating LLMs on vulnerability detection~\cite{zhou2024large,gao2023far,steenhoek2024comprehensive,sp2024,ding2024vulnerability}. Table~\ref{tab:related-work-comparison} provides a comparison against these studies. Section~\ref{sec:rq1} corroborates the common finding from these studies
that LLMs do not perform well on real-world datasets. However, to the best of our knowledge,
our study is the first work that qualitatively identifies prompts and vulnerability classes where LLMs perform well (and often better than other tools). Moreover, our study focuses on a larger/different class of state-of-the-art LLMs, datasets, languages, and vulnerability classes.

 \section{Threats to Validity}
\mypara{External} The choice of LLMs and datasets may bias our evaluation and insights. To address
this threat, we choose multiple popular synthetic and real-world datasets across two
languages: Java and C++. We also choose both state-of-the-art closed-source and
open-source LLMs. \Comment{Hence, our study is comprehensive.}However, our insights
may not generalize to other languages or datasets.

\mypara{Internal} Owing to the non-deterministic nature of LLMs 
and single experiment runs per benchmark,
our observations may be biased. To mitigate this threat, 
we use a temperature of 0 to ensure determinism across all LLMs. While this works well for 
locally run CodeLlama models, it is well-known that GPT-4 and GPT-3.5 might still return 
non-deterministic results. However, 
this should balance out across datasets and over the
large number of benchmarks we evaluate on.
Further, given similar results across LLMs on real-world-datasets, 
we do not expect
significant changes with re-runs.

Our evaluation code and scripts may have bugs, which might bias our results. 
Our manual analysis of results may lead to erroneous inferences. To
address this threat, multiple co-authors reviewed the code regularly and
actively fixed issues.  Further, multiple co-authors independently  
analyzed the results and discussed them together to mitigate any discrepancies.

\Comment{Our choice of prior vulnerability detection tools may not be complete. To
address this threat, we compare the LLMs against one of the most popular and
widely-used static analysis tool, CodeQL, as well as a state-of-the-art deep
learning-based tool (LineVul) that has previously demonstrated the best
vulnerability detection performance. }

 \section{Conclusion}
In this work, we performed a comprehensive analysis of LLMs for security
vulnerability detection. Our study reveals that both closed-source LLMs, such as
GPT-4, and open-source LLMs, like \clama, perform modestly at vulnerability detection for both Java and C/C++.
However, we find specific vulnerability classes where LLMs excel (often performing better than static analysis tools, such as CodeQL).
Moreover, we find that even in cases where the models produce incorrect predictions, they are capable of identifying relevant sources, sinks
and sanitizers that can be used for downstream dataflow analysis.
Hence, we believe that an interesting future direction is to develop neuro-symbolic techniques that combine
the intuitive reasoning abilities of LLMs with symbolic tools such as logical reasoning engines and static code 
analyzers for more effective and interpretable solutions. 
\Comment{We share all our code, experiment scripts, and the appendix in the supplementary material. We will open-source a
replication package on acceptance.}

\clearpage

\bibliographystyle{IEEEtran}
\bibliography{references}

\clearpage
\newpage 
\section{Appendix}
\subsection{Dataset Processing and Selection}
\label{sec:datasel}
We perform a data processing and cleaning step for each dataset before
evaluating them with LLMs. 

\mypara{OWASP} We remove or anonymize information in OWASP benchmarks that may
provide obvious hints about the vulnerability in a file. For instance, we change
package, variable names, and strings such as ``owasp'', ``testcode'', and
``/sqli-06/BenchmarkTest02732'' to other pre-selected un-identifying names such
as ``pcks'', ``csdr'', etc. We remove all comments in the file because they may
explicitly highlight the vulnerable line of code or may have irrelevant text
(such as copyright info), which may leak information. These changes, however, do
not change the semantics of the code snippets.

\mypara{Juliet Java and C/C++} Similar to OWASP, we remove all comments and
transform all identifiers that leak identifying information in all test cases in
the Juliet benchmark. For instance, we change ``class
CWE80\_XSS\_CWE182\_Servlet\_connect\_tcp\_01'' to ``class MyClass''. The
Juliet benchmark provides the vulnerable (named as ``bad'') and non-vulnerable
(named as ``good*'') methods in the same file. For easier evaluation, we perform
a pre-processing step to split each file into two, each containing either a
vulnerable or non-vulnerable method. Juliet also contains special benchmarks
that have dependencies across multiple (2-5) files. We skip these benchmarks
because they are typically too big to fit into the LLM prompt. Hence, the
number of test cases after the data processing step in Juliet is reduced 
(as shown in Table~\ref{tab:processing}).

\mypara{CVEFixes} For each CVE, CVEFixes provides the methods that were involved
in the fix commit. It also includes the method code in the parent commit, i.e.,
the method version before the fix. We collect all methods in the fix commit and
the parent commit and label them as vulnerable and non-vulnerable, respectively.
Similar to other datasets, we also remove all comments in the method code. While
CVEFixes contains methods across multiple programming languages, we only collect
C/C++ and Java methods for our study.

\subsection{Prompting Strategies}
\label{sec:app:prompts}
The \basic is presented in Listing~\ref{lst:promptformatbasic},
\cwespecificstr in Listing~\ref{lst:promptformatcwespecific}
and \dfstr in Listing~\ref{lst:promptformatdf}.
\begin{figure}[!htb]
\centering
\begin{lstlisting}[style=mypromptstyle,caption={Basic LLM Prompt},label={lst:promptformatbasic},numbers=none,escapechar=@]
System: You are a security researcher, expert in detecting security vulnerabilities. 
Provide response only in following format: vulnerability: <YES or NO> | vulnerability type: <CWE ID> | vulnerability name: <CWE NAME> | explanation: <explanation for prediction>. 
Use N/A in other fields if there are no vulnerabilities. Do not include anything else in response.

User: Is the following code snippet prone to any security vulnerability?
<CODE_SNIPPET>
Response: 
\end{lstlisting}
\vspace{-0.1in}
\end{figure}

\begin{figure}[!htb]
\centering
\begin{lstlisting}[style=mypromptstyle,caption={CWE-specific LLM Prompt},label={lst:promptformatcwespecific},numbers=none,escapechar=@]
System: [Same as above]
User: Is the following code snippet prone to <CWE>?
<CODE_SNIPPET>
Response: 
\end{lstlisting}
\vspace{-0.1in}
\end{figure}

\begin{figure}[!ht]
\begin{lstlisting}[style=mypromptstyle,caption={Dataflow analysis-based LLM Prompt},label={lst:promptformatdf},numbers=none]
System: You are a security researcher, expert in detecting security vulnerabilities. Carefully analyze the given code snippet and track the data flows from various sources to sinks. Assume that any call to an unknown external API is unsanitized.

Please provide a response only in the following itemized OUTPUT FORMAT. Use N/A in other fields if there are no vulnerabilities. DO NOT INCLUDE ANYTHING ELSE IN YOUR RESPONSE.
<OUTPUT FORMAT>
Data flow analysis of the given code snippet:
1. Sources: 
<numbered list of input sources>
2. Sinks:
<numbered list of output sinks>
3. Sanitizers:
<numbered list of sanitizers, if any>
4. Unsanitized Data Flows:
<numbered list of data flows that are not sanitized in the format (source, sink, why this flow could be vulnerable)>
5. Final Vulnerability analysis verdict: vulnerability: <YES or NO> | vulnerability type: <CWE_ID> | vulnerability name: <NAME_OF_CWE> | explanation: <explanation for prediction> 
</OUTPUT FORMAT>
User: Is the following code snippet prone to <CWE>?
<CODE_SNIPPET>
Response:
\end{lstlisting}
\vspace{-0.1in}
\end{figure}

\subsection{Other Prompting Strategies}
\label{sec:otherprompting}
In addition to the prompting strategies presented in our main evaluation, we considered other popular prompting strategies such as \texttt{Few-shot prompting} and \texttt{Chain-of-thought prompting} in a limited experimental setting. For the few-shot prompt (\fewshot), we included two examples of the task (one with a vulnerability and one without) in the \cwespecificstr before providing the target code snippet. For the chain-of-thought prompt (\chot), we explicitly ask the model to provide a reasoning chain before the final answer by adding a ``Let's think step-by-step'' statement at the end of the \cwespecificstr. The \chot and \fewshot prompts are provided in Listing~\ref{lst:promptformatcot} and Listing~\ref{lst:promptformatfs} respectively.

\begin{figure}[!htb]
\centering
\begin{lstlisting}[style=mypromptstyle,caption={\chot LLM Prompt},label={lst:promptformatcot},numbers=none,escapechar=@,inputencoding=utf8]
System: [Same as the Basic prompt]
User: Is the following code snippet prone to <CWE>?
<CODE_SNIPPET>
Let's think step by step.
Response: 
\end{lstlisting}
\vspace{-0.1in}
\end{figure}

\begin{figure}[!htb]
\centering
\begin{lstlisting}[style=mypromptstyle,caption={\fewshot LLM Prompt},label={lst:promptformatfs},numbers=none,escapechar=@,inputencoding=utf8]
System: [Same as the Basic prompt]
User: 

Query: Is the following code snippet prone to <CWE1>?
Code snippet: <CODE_SNIPPET1>

Vulnerability analysis verdict: $ vulnerability: YES | vulnerability type: <CWE1> . . .

Query: Is the following code snippet prone to <CWE2>?
Code snippet: <CODE_SNIPPET2>

Vulnerability analysis verdict: $ vulnerability: NO | vulnerability type: N/A . . .

Query: Is the following code snippet prone to <CWE>?
Code snippet: <CODE_SNIPPET>

Vulnerability analysis verdict: 
\end{lstlisting}
\vspace{-0.1in}
\end{figure}

Table~\ref{tab:gpt4_julietj_all_prompts} and Table~\ref{tab:gpt4_cvec_all_prompts} present the results from GPT-4 with various prompting strategies on a random subset of 100 samples of the \julietj and \cvec datasets respectively. The \df prompt reports the highest accuracy of 69\% and the highest F1 score of 0.75 on the \julietj dataset. The \df prompt reports a 0.05 higher F1 score than the \chot prompt and a 0.03 higher F1 score than the \fewshot prompt. This difference is much more prominent on the \cvec dataset where the \df prompt reports a 0.34 higher F1 score than the \chot prompt and a 0.31 higher F1 score than the \fewshot prompt. Moreover, the \fewshot prompt reported a 0.2 lower F1 score than the \cwespecificstr on the \cvec dataset while requiring more tokens. Our analysis of the few-shot prompts suggests that providing more examples may not be a useful strategy for vulnerability detection. Because the potential set of vulnerable code patterns is quite large, the provided examples hardly make a difference to LLMs' reasoning abilities. Hence, it may be more useful to use prompts that instead elicit reasoning or explanations of some kind before detecting if the given snippet is vulnerable. The \chot prompt, however, does not help with reasoning always, as it either performed at par or worse than the \dfstr. 

\begin{table}[!hbt]
    \centering
    \footnotesize
     \caption{All prompting strategies on 100 samples from \julietj.}
    \label{tab:gpt4_julietj_all_prompts}
    \vspace{-0.1in}
    \setlength{\tabcolsep}{0.5em} 
\begin{tabular}{llHllll}
\midrule
\textbf{Model}   & \textbf{Prompt}   & \textbf{\julietj}   &\multicolumn{4}{c}{\textbf{Metrics}}             \\
\midrule
       &          & C        & A           & P    & R    & F1   \\ \midrule
 GPT-4 & \cwespecific & 100      & 0.65          & 0.58 & 0.96 & 0.72 \\
 GPT-4 & \fewshot     & 100      & 0.65          & 0.58 & 0.94 & 0.72 \\
 GPT-4 & \chot        & 100      & 0.69          & 0.64 & 0.79 & 0.70 \\
 GPT-4 & \df          & 100      & \textbf{0.69}          & 0.61 & 0.96 & \textbf{0.75} \\
\bottomrule
\end{tabular}
\end{table}

\begin{table}[!hbt]
    \centering
    \footnotesize
     \caption{All prompting strategies on 100 samples from \cvec.}
    \label{tab:gpt4_cvec_all_prompts}
    \vspace{-0.1in}
    \setlength{\tabcolsep}{0.5em} 
\begin{tabular}{llHrrrrH}
\toprule
\textbf{Model}   & \textbf{Prompt}   & \textbf{\cvec}   & \multicolumn{4}{c}{\textbf{Metrics}}    &       \\
\midrule
       &              & C       & A      & P                & R    & F1   & Time  \\ \midrule
 GPT-4 & \cwespecific & 97      & 0.55     & 0.54             & 0.58 & 0.56 & 25.22 \\
 GPT-4 & \fewshot     & 99      & 0.49     & 0.38             & 0.34 & 0.36 & 15.10 \\
 GPT-4 & \chot         & 93      & 0.52     & 0.37             & 0.30 & 0.33 & 26.16 \\
 GPT-4 & \df          & 98      & \textbf{0.56}     & 0.56             & 0.83 & \textbf{0.67} & 29.93 \\
\bottomrule
\end{tabular}
\end{table}


Learning from these experiments, we selected the \cwespecificstr, \dfstr, in addition to the \generic prompt, for our main evaluation with LLMs.


\subsection{Detailed metrics across all LLMs and Datasets}
\label{sec:detailedmetrics}
Table~\ref{tab:llms_performance_all} presents the metrics for all LLMs and Datasets across all prompts.
\begin{table*}[!htb]
    \footnotesize
    \centering
    \setlength{\tabcolsep}{0.4em}
  \caption{ Effectiveness of LLMs in Predicting Security Vulnerabilities (Java and C++). The highest accuracy and F1 scores (as well as ones within 0.1 range of the highest values) for each dataset are highlighted in blue.}
    \label{tab:llms_performance_all}
        \begin{tabular}{ll|Hrrrr|Hrrrr|Hrrrr||Hrrrr|Hrrrr}
            \toprule
             \textbf{Model}         & \textbf{Prompt}       & \multicolumn{5}{|c|}{\textbf{OWASP}} & \multicolumn{5}{|c|}{\textbf{Juliet Java}} & \multicolumn{5}{|c|}{\textbf{CVEFixes Java}} & \multicolumn{5}{|c|}{\textbf{Juliet C/C++}} & \multicolumn{5}{|c}{\textbf{CVEFixes C/C++}}  \\\midrule
                           &              & C    & \textbf{A}     & \textbf{P}        & \textbf{R}       & \textbf{F1}   & C    & \textbf{A}     & \textbf{P}        & \textbf{R}       & \textbf{F1}    & C    & \textbf{A}     & \textbf{P}        & \textbf{R}       & \textbf{F1} & C    & \textbf{A}     & \textbf{P}        & \textbf{R}       & \textbf{F1}   & C    & \textbf{A} & \textbf{P}             & \textbf{R}    & \textbf{F1}    \\\midrule
                         \qwenc & \generic     & 1000 & 0.50          & 0.50 & 0.82    & 0.62          & 1000    & 0.50          & 0.50 & 0.99 & 0.66          & 1000 & 0.49          & 0.49 & 0.68            & 0.57          & 1000 & 0.49          & 0.50           & 0.99 & 0.66          & 1000 & 0.51             & 0.51 & 0.78 & 0.61          \\
                         \qwenc & \cwespecific & 1000 & 0.49          & 0.49 & 0.79    & 0.61          & 1000    & 0.50          & 0.50 & 1.00 & 0.67          & 1000 & 0.51          & 0.50 & 0.92            & 0.65          & 1000 & 0.50          & 0.50           & 1.00 & 0.67          & 1000 & 0.51             & 0.50 & 0.89 & 0.64          \\
                         \qwenc & \df          & 1000 & 0.47          & 0.48 & 0.75    & 0.59          & 1000    & 0.55          & 0.54 & 0.67 & 0.60          & 1000 & 0.50          & 0.50 & 0.80            & 0.62          & 1000 & 0.57          & 0.55           & 0.79 & 0.65          & 1000 & 0.52             & 0.51 & 0.77 & 0.62          \\ \midrule
                         \qwencseven   & \generic     & 1000 & 0.50          & 0.50 & 1.00    & 0.67          & 1000    & 0.50          & 0.50 & 0.99 & 0.67          & 1000 & 0.47          & 0.48 & 0.79            & 0.60          & 1000 & 0.50          & 0.50           & 1.00 & 0.67          & 1000 & 0.50             & 0.50 & 0.95 & \cellhl{0.66} \\
                         \qwencseven   & \cwespecific & 1000 & 0.50          & 0.50 & 1.00    & 0.67          & 1000    & 0.50          & 0.50 & 1.00 & 0.67          & 1000 & 0.48          & 0.49 & 0.53            & 0.51          & 1000 & 0.50          & 0.50           & 1.00 & 0.66          & 1000 & 0.51             & 0.50 & 0.77 & 0.61          \\
                         \qwencseven   & \df          & 1000 & 0.54          & 0.52 & 1.00    & 0.68          & 1000    & 0.52          & 0.51 & 0.99 & 0.67          & 1000 & 0.52          & 0.52 & 0.49            & 0.50          & 1000 & 0.50          & 0.50           & 0.99 & 0.67          & 1000 & 0.54             & 0.53 & 0.62 & 0.57          \\ \midrule
                         \clamasvnabv        & \generic     & 1000 & 0.51          & 0.87 & 0.03    & 0.05          & 1000    & 0.51          & 0.59 & 0.09 & 0.15          & 1000 & 0.47          & 0.29 & 0.04            & 0.06          & 1000 & 0.50          & 0.50           & 0.12 & 0.19          & 1000 & 0.49             & 0.33 & 0.02 & 0.03          \\
                         \clamasvnabv        & \cwespecific & 1000 & 0.50          & 0.50 & 1.00    & 0.67          & 1000    & 0.52          & 0.51 & 0.99 & 0.67          & 1000 & 0.51          & 0.51 & 0.84            & 0.63          & 1000 & 0.51          & 0.50           & 0.99 & 0.67          & 1000 & 0.50             & 0.50 & 0.85 & 0.63          \\
                         \clamasvnabv        & \df          & 1000 & 0.50          & 0.50 & 1.00    & 0.67          & 1000    & 0.50          & 0.50 & 1.00 & 0.67          & 1000 & 0.50          & 0.50 & 1.00            & \cellhl{0.67} & 1000 & 0.50          & 0.50           & 1.00 & 0.67          & 1000 & 0.51             & 0.50 & 0.97 & \cellhl{0.66} \\ \midrule
                         \dscoderseven          & \generic     & 1000 & 0.50          & 0.50 & 0.99    & 0.66          & 1000    & 0.57          & 0.56 & 0.69 & 0.62          & 1000 & 0.48          & 0.47 & 0.30            & 0.36          & 1000 & 0.57          & 0.55           & 0.77 & 0.64          & 1000 & 0.49             & 0.47 & 0.24 & 0.32          \\
                         \dscoderseven          & \cwespecific & 1000 & 0.56          & 0.54 & 0.87    & 0.66          & 1000    & 0.54          & 0.53 & 0.75 & 0.62          & 1000 & 0.48          & 0.43 & 0.15            & 0.22          & 1000 & 0.58          & 0.56           & 0.70 & 0.62          & 1000 & 0.51             & 0.53 & 0.18 & 0.27          \\
                         \dscoderseven          & \df          & 1000 & 0.51          & 0.50 & 0.98    & 0.66          & 1000    & 0.52          & 0.51 & 0.91 & 0.65          & 1000 & 0.49          & 0.50 & 0.90            & 0.64          & 1000 & 0.50          & 0.50           & 0.98 & 0.66          & 1000 & 0.53             & 0.52 & 0.90 & \cellhl{0.66} \\ \midrule
                         \llamaeight        & \generic     & 1000 & 0.50          & 0.50 & 1.00    & 0.67          & 1000    & 0.48          & 0.49 & 0.94 & 0.65          & 1000 & 0.52          & 0.51 & 0.80            & 0.62          & 1000 & 0.49          & 0.49           & 0.97 & 0.65          & 1000 & 0.52             & 0.51 & 0.92 & \cellhl{0.66} \\
                         \llamaeight        & \cwespecific & 1000 & 0.53          & 0.52 & 1.00    & 0.68          & 1000    & 0.52          & 0.51 & 0.97 & 0.67          & 1000 & 0.53          & 0.56 & 0.29            & 0.38          & 1000 & 0.54          & 0.52           & 0.98 & 0.68          & 1000 & \cellhl{0.55}    & 0.55 & 0.58 & 0.56          \\
                         \llamaeight        & \df          & 1000 & 0.49          & 0.50 & 0.93    & 0.65          & 1000    & 0.50          & 0.50 & 0.97 & 0.66          & 1000 & 0.51          & 0.50 & 0.93            & 0.65          & 1000 & 0.50          & 0.50           & 0.99 & 0.67          & 1000 & 0.50             & 0.50 & 0.95 & 0.65          \\ \midrule
                         \clamathrabv        & \generic     & 1000 & \cellhl{0.60} & 0.58 & 0.74    & 0.65          & 1000    & 0.48          & 0.48 & 0.41 & 0.44          & 1000 & 0.50          & 0.51 & 0.08            & 0.14          & 1000 & 0.47          & 0.47           & 0.51 & 0.49          & 1000 & 0.50             & 0.50 & 0.07 & 0.12          \\
                         \clamathrabv        & \cwespecific & 1000 & 0.52          & 0.51 & 0.98    & 0.67          & 1000    & 0.50          & 0.50 & 0.89 & 0.64          & 1000 & 0.48          & 0.47 & 0.29            & 0.36          & 1000 & 0.53          & 0.51           & 0.98 & 0.67          & 1000 & 0.53             & 0.52 & 0.56 & 0.54          \\
                         \clamathrabv        & \df          & 1000 & 0.50          & 0.50 & 1.00    & 0.67          & 1000    & 0.50          & 0.50 & 1.00 & 0.67          & 1000 & 0.50          & 0.50 & 1.00            & \cellhl{0.67} & 1000 & 0.50          & 0.50           & 1.00 & 0.67          & 1000 & 0.50             & 0.50 & 0.96 & \cellhl{0.66} \\ \midrule
                         \qwenfourteen        & \generic     & 1000 & 0.54          & 0.52 & 1.00    & 0.68          & 1000    & 0.50          & 0.50 & 0.74 & 0.60          & 1000 & 0.53          & 0.54 & 0.43            & 0.48          & 1000 & 0.48          & 0.49           & 0.74 & 0.59          & 1000 & 0.52             & 0.52 & 0.53 & 0.52          \\
                         \qwenfourteen        & \cwespecific & 1000 & 0.57          & 0.54 & 0.92    & 0.68          & 1000    & 0.71          & 0.65 & 0.87 & 0.75          & 1000 & 0.55          & 0.62 & 0.25            & 0.36          & 1000 & \cellhl{0.65} & 0.60           & 0.89 & 0.72          & 1000 & 0.52             & 0.52 & 0.32 & 0.39          \\
                         \qwenfourteen        & \df          & 1000 & 0.55          & 0.52 & 1.00    & 0.69          & 1000    & 0.66          & 0.61 & 0.88 & 0.72          & 1000 & 0.56          & 0.58 & 0.42            & 0.49          & 1000 & \cellhl{0.64} & 0.59           & 0.95 & \cellhl{0.73} & 1000 & \cellhl{0.55}    & 0.56 & 0.45 & 0.50          \\ \midrule
                         \dscoderfifteen      & \generic     & 1000 & 0.50          & 0.50 & 1.00    & 0.67          & 1000    & 0.54          & 0.52 & 0.97 & 0.68          & 1000 & 0.44          & 0.44 & 0.44            & 0.44          & 1000 & 0.51          & 0.50           & 0.98 & 0.67          & 1000 & 0.49             & 0.49 & 0.26 & 0.34          \\
                         \dscoderfifteen      & \cwespecific & 1000 & 0.50          & 0.50 & 1.00    & 0.67          & 1000    & 0.50          & 0.50 & 1.00 & 0.67          & 1000 & 0.52          & 0.51 & 0.93            & \cellhl{0.66} & 1000 & 0.50          & 0.50           & 1.00 & 0.67          & 1000 & 0.50             & 0.50 & 0.95 & \cellhl{0.66} \\
                         \dscoderfifteen      & \df          & 1000 & 0.50          & 0.50 & 1.00    & 0.67          & 1000    & 0.50          & 0.50 & 1.00 & 0.67          & 1000 & 0.51          & 0.51 & 0.86            & 0.64          & 1000 & 0.50          & 0.50           & 1.00 & 0.67          & 1000 & 0.51             & 0.50 & 0.94 & 0.66          \\ \midrule
                         codestral-22b       & \generic     & 1000 & 0.50          & 0.50 & 1.00    & 0.67          & 1000    & 0.52          & 0.51 & 0.91 & 0.65          & 1000 & 0.49          & 0.49 & 0.63            & 0.55          & 1000 & 0.50          & 0.50           & 0.93 & 0.65          & 1000 & 0.50             & 0.50 & 0.40 & 0.44          \\
                         codestral-22b       & \cwespecific & 1000 & 0.52          & 0.51 & 0.98    & 0.67          & 1000    & 0.52          & 0.51 & 0.96 & 0.67          & 1000 & 0.50          & 0.50 & 0.37            & 0.43          & 1000 & 0.57          & 0.54           & 0.93 & 0.69          & 1000 & 0.52             & 0.56 & 0.16 & 0.25          \\
                         codestral-22b       & \df          & 1000 & 0.53          & 0.52 & 1.00    & 0.68          & 1000    & 0.50          & 0.50 & 0.99 & 0.67          & 1000 & 0.53          & 0.52 & 0.89            & 0.65          & 1000 & 0.50          & 0.50           & 0.99 & 0.67          & 1000 & 0.52             & 0.51 & 0.87 & 0.64          \\ \midrule
                         \qwenthirtytwo        & \generic     & 1000 & 0.52          & 0.51 & 1.00    & 0.67          & 1000    & 0.48          & 0.49 & 0.77 & 0.60          & 1000 & 0.52          & 0.53 & 0.38            & 0.44          & 1000 & 0.50          & 0.50           & 0.84 & 0.63          & 1000 & 0.47             & 0.46 & 0.36 & 0.41          \\
                         \qwenthirtytwo        & \cwespecific & 1000 & 0.56          & 0.53 & 1.00    & \cellhl{0.69} & 1000    & 0.58          & 0.55 & 0.93 & 0.69          & 1000 & 0.53          & 0.55 & 0.30            & 0.39          & 1000 & 0.63          & 0.58           & 0.87 & 0.70          & 1000 & 0.53             & 0.54 & 0.35 & 0.43          \\
                         \qwenthirtytwo        & \df          & 1000 & 0.55          & 0.52 & 1.00    & 0.69          & 1000    & 0.59          & 0.55 & 1.00 & 0.71          & 1000 & 0.55          & 0.54 & 0.62            & 0.58          & 1000 & 0.54          & 0.52           & 0.98 & 0.68          & 1000 & \cellhl{0.56}    & 0.54 & 0.81 & 0.65          \\ \midrule
                         \dscoderthirtythree         & \generic     & 1000 & 0.52          & 0.51 & 0.97    & 0.67          & 1000    & 0.56          & 0.53 & 0.94 & 0.68          & 1000 & 0.50          & 0.50 & 0.60            & 0.55          & 1000 & 0.42          & 0.46           & 0.81 & 0.58          & 1000 & 0.51             & 0.51 & 0.75 & 0.60          \\
                         \dscoderthirtythree         & \cwespecific & 1000 & 0.53          & 0.52 & 0.86    & 0.65          & 1000    & 0.56          & 0.54 & 0.85 & 0.66          & 1000 & 0.49          & 0.49 & 0.39            & 0.43          & 1000 & 0.44          & 0.46           & 0.78 & 0.58          & 1000 & 0.52             & 0.52 & 0.56 & 0.54          \\
                         \dscoderthirtythree         & \df          & 1000 & 0.51          & 0.51 & 0.75    & 0.60          & 1000    & 0.46          & 0.47 & 0.63 & 0.54          & 1000 & 0.53          & 0.53 & 0.64            & 0.58          & 1000 & 0.50          & 0.50           & 0.78 & 0.61          & 1000 & 0.49             & 0.49 & 0.54 & 0.52          \\ \midrule
                         \clamathirtyfourabv & \generic     & 1000 & 0.51          & 0.50 & 1.00    & 0.67          & 1000    & 0.47          & 0.48 & 0.85 & 0.62          & 1000 & 0.50          & 0.50 & 0.28            & 0.36          & 1000 & 0.50          & 0.50           & 0.93 & 0.65          & 1000 & 0.51             & 0.52 & 0.20 & 0.29          \\
                         \clamathirtyfourabv & \cwespecific & 1000 & 0.57          & 0.54 & 0.94    & 0.69          & 1000    & 0.49          & 0.49 & 0.94 & 0.65          & 1000 & 0.50          & 0.51 & 0.17            & 0.25          & 1000 & 0.53          & 0.52           & 0.98 & 0.68          & 1000 & 0.51             & 0.54 & 0.08 & 0.14          \\
                         \clamathirtyfourabv & \df          & 1000 & 0.50          & 0.50 & 1.00    & 0.67          & 1000    & 0.50          & 0.50 & 1.00 & 0.67          & 1000 & 0.50          & 0.50 & 1.00            & \cellhl{0.67} & 1000 & 0.50          & 0.50           & 1.00 & 0.67          & 1000 & 0.50             & 0.50 & 0.99 & \cellhl{0.67} \\ \midrule
                         \llamaseventy       & \generic     & 1000 & 0.51          & 0.50 & 1.00    & 0.67          & 1000    & 0.51          & 0.51 & 0.84 & 0.63          & 1000 & 0.51          & 0.51 & 0.71            & 0.59          & 1000 & 0.53          & 0.52           & 0.92 & 0.66          & 1000 & 0.51             & 0.51 & 0.90 & 0.65          \\
                         \llamaseventy       & \cwespecific & 1000 & 0.58          & 0.54 & 0.99    & \cellhl{0.70} & 1000    & \cellhl{0.76} & 0.71 & 0.89 & \cellhl{0.79} & 1000 & 0.52          & 0.53 & 0.43            & 0.48          & 1000 & 0.59          & 0.55           & 0.95 & 0.70          & 1000 & 0.52             & 0.51 & 0.71 & 0.60          \\
                         \llamaseventy       & \df          & 1000 & 0.54          & 0.52 & 0.99    & 0.68          & 1000    & 0.72          & 0.68 & 0.84 & 0.75          & 1000 & 0.55          & 0.54 & 0.63            & 0.58          & 1000 & 0.59          & 0.55           & 0.96 & 0.70          & 1000 & 0.54             & 0.53 & 0.77 & 0.63          \\ \midrule
                         \gemini    & \generic     & 1000 & 0.54          & 0.52 & 0.98    & 0.68          & 1000    & 0.47          & 0.48 & 0.76 & 0.59          & 1000 & 0.52          & 0.52 & 0.53            & 0.52          & 1000 & 0.44          & 0.46           & 0.81 & 0.59          & 1000 & 0.47             & 0.47 & 0.51 & 0.49          \\
                         \gemini    & \cwespecific & 1000 & 0.57          & 0.54 & 1.00    & \cellhl{0.70} & 1000    & 0.51          & 0.51 & 0.91 & 0.65          & 1000 & 0.54          & 0.57 & 0.31            & 0.40          & 1000 & 0.50          & 0.50           & 0.89 & 0.64          & 1000 & 0.51             & 0.51 & 0.52 & 0.51          \\
                         \gemini    & \df          & 1000 & 0.54          & 0.52 & 1.00    & 0.68          & 1000    & 0.50          & 0.50 & 1.00 & 0.67          & 1000 & \cellhl{0.57} & 0.55 & 0.79            & 0.65          & 1000 & 0.50          & 0.50           & 0.99 & 0.66          & 1000 & 0.51             & 0.50 & 0.86 & 0.64          \\ \midrule
                         GPT-3.5             & \generic     & 1000 & 0.52          & 0.52 & 0.72    & 0.60          & 1000    & 0.58          & 0.57 & 0.71 & 0.63          & 1000 & 0.46          & 0.35 & 0.09            & 0.15          & 1000 & 0.49          & 0.49           & 0.64 & 0.56          & 1000 & 0.52             & 0.56 & 0.20 & 0.29          \\
                         GPT-3.5             & \cwespecific & 1000 & 0.55          & 0.54 & 0.62    & 0.58          & 1000    & 0.52          & 0.52 & 0.55 & 0.54          & 1000 & 0.47          & 0.41 & 0.12            & 0.19          & 1000 & 0.49          & 0.49           & 0.70 & 0.58          & 1000 & 0.52             & 0.54 & 0.19 & 0.28          \\
                         GPT-3.5             & \df          & 1000 & 0.51          & 0.50 & 0.93    & 0.65          & 1000    & 0.40          & 0.44 & 0.73 & 0.55          & 1000 & 0.54          & 0.53 & 0.66            & 0.59          & 1000 & 0.40          & 0.44           & 0.77 & 0.56          & 1000 & 0.52             & 0.52 & 0.75 & 0.61          \\ \midrule
                         GPT-4               & \generic     & 1000 & 0.52          & 0.51 & 1.00    & 0.67          & 1000    & 0.56          & 0.54 & 0.85 & 0.66          & 1000 & 0.50          & 0.50 & 0.34            & 0.41          & 1000 & 0.54          & 0.52           & 0.92 & 0.67          & 1000 & 0.51             & 0.51 & 0.57 & 0.54          \\
                         GPT-4               & \cwespecific & 1000 & 0.54          & 0.52 & 1.00    & 0.68          & 1000    & 0.69          & 0.63 & 0.97 & 0.76          & 1000 & 0.55          & 0.56 & 0.44            & 0.49          & 1000 & 0.58          & 0.54           & 0.95 & 0.69          & 1000 & 0.52             & 0.52 & 0.52 & 0.52          \\
                         GPT-4               & \df          & 1000 & 0.55          & 0.52 & 1.00    & 0.69          & 1000    & 0.70          & 0.63 & 0.98 & 0.76          & 1000 & 0.53          & 0.53 & 0.59            & 0.56          & 1000 & 0.59          & 0.55           & 0.98 & 0.70          & 1000 & 0.52             & 0.51 & 0.76 & 0.61          \\
                           \bottomrule
        \end{tabular}
    \end{table*}

\Comment{
\subsection{Runtime statistics of LLMs}
In this section, we discuss the average time taken per query for different models and prompting setups. In general, the GPT-4 OpenAI API calls are faster than running inference with CodeLlama locally. The \df prompt takes 21.08 seconds with GPT-4 and 89.29 seconds and 82.07 seconds, respectively, with the CodeLlama-13B and CodeLlama-7B models. The self-reflection step additionally takes 10.28 seconds with GPT-4. The \generic and \cwespecific prompts are faster to run, taking 9.32 seconds and 14.64 seconds with GPT-4, respectively.

\begin{table}[!htb]
    \centering
    \footnotesize
     \caption{Average Runtime statistics}
    \label{tab:runtime_statistics}
    \vspace{-0.1in}
    \setlength{\tabcolsep}{0.5em} 
\begin{tabular}{llr}
\toprule
\textbf{Model}   & \textbf{Prompt}   & \textbf{Average time taken per query (secs)} \\
\midrule
GPT-4 & \generic & 9.32  \\
 GPT-4 & \cwespecific & 14.64  \\
 GPT-4 & \df          & 21.08 \\
\midrule
CodeLlama-13B & \generic & 25.56  \\
CodeLlama-13B & \cwespecific & 24.82  \\
CodeLlama-13B & \df          & 89.29 \\
\midrule
CodeLlama-7B & \generic & 13.98  \\
CodeLlama-7B & \cwespecific & 15.05 \\
CodeLlama-7B & \df          & 82.07 \\
\bottomrule
\end{tabular}
\end{table}
}

\subsection{More examples of CodeQL vs. LLMs}
\label{appendix:codeql}

\begin{figure}[!htb]
\begin{minted}{c}
// Code snippet prone to CWE-476 (NULL Pointer Dereference)
// Both the conditions are evaluated due to the bitwise & operator

if(|\colorbox{yellow}{globalReturnsTrue()}|)
{
    twoIntsStruct |\colorbox{yellow}{*twoIntsStructPointer}| = NULL;
    if (|\colorbox{pink}{(twoIntsStructPointer != NULL)}| 
        & 
        |\colorbox{pink}{(twoIntsStructPointer->intOne == 5)}|)
    {
        printLine("intOne == 5");
    }
}


\end{minted}
\vspace{-0.1in}
\prediction{The pointer \texttt{twoIntsStructPointer} is initialized as \texttt{NULL} and then checked if it's not \texttt{NULL} before dereferencing. However, the pointer is not assigned any value other than \texttt{NULL}, so it's not actually a risk of NULL pointer dereference. \xmark}
\vspace{-0.1in}
\captionof{lstlisting}{CodeQL correctly predicts Null Pointer Dereferencing while GPT-4 doesn't. 
GPT-4, however, correctly identifies the \colorbox{yellow}{sources} and \colorbox{pink}{sinks}.}
\label{fig:julietcpp47137}
\end{figure}

In Listing~\ref{fig:julietcpp47137}, the model incorrectly reasons about the sanitization by
overlooking the \& which would cause both the expressions to be evaluated.
The identified sources and sinks are correct, however.

\subsection{Qualitative analysis of GPT-4 responses}
\label{appendix:gpt4_cwe_df}

We first present examples where the dataflow analysis from the \df prompt is
useful. Consider the code snippet in
Listing~\ref{fig:cvefixesjava3470}. In this snippet, the variable \texttt{dir} is indirectly being used to create a directory via the \texttt{dirToCreate} variable. \emph{GPT-4 correctly identifies that this path is not sanitized and could be used to create a directory in otherwise restricted locations}. This could lead to CWE-22 (path traversal) as is rightly predicted by GPT-4. In Listing~\ref{fig:cvefixesjava2389}, the method \texttt{write} takes an array \texttt{b} and integers \texttt{offset} and \texttt{length} as inputs. \emph{The expression \texttt{offset + length > b.length} can lead to an Integer Overflow (CWE-190) if \texttt{offset + length} exceeds the maximum integer value} and this is correctly identified by GPT-4. Finally, in Listing~\ref{fig:cvefixesjava2241}, there are multiple array accesses that could lead to CWE-125 (Out-of-bounds read). \emph{GPT-4 correctly identifies that the expression \texttt{ciphertextOffset + length + index} could exceed the size of the \texttt{ciphertext} array}, leading to this vulnerability. 

\setlength{\fboxsep}{0pt}
\begin{figure}[!htb]
\begin{lstlisting}[style=java,caption={Correct response by GPT-4 on a Path Traversal vulnerability},label={fig:cvefixesjava3470},numbers=none,framesep=2pt,frame=lrtb,framerule=1pt,escapechar=@]
// CVE-2021-39180

// CODE SNIPPET

public FileVisitResult preVisitDirectory(Path dir, BasicFileAttributes attrs) throws IOException {
    Path relativeDir = source.relativize(dir);
    final Path dirToCreate = Paths.get(destDir.toString(), relativeDir.toString());
    if(!dirToCreate.toFile().exists()) {
        Files.createDirectory(dirToCreate);
    }
    return FileVisitResult.CONTINUE;
}

// RESPONSE WITH DATAFLOW ANALYSIS INSPIRED PROMPT
Here is a data flow analysis of the given code snippet:
A. Sources: 
1. The `dir` parameter in the `preVisitDirectory` method.
2. The `source` variable.
3. The `destDir` variable.

B. Sinks:
1. The `Files.createDirectory` method.

C. Sanitizers:
No sanitizers are present in the code snippet.

D. Unsanitized Data Flows:
1. (dir, Files.createDirectory, The `dir` parameter is used to create a new directory without any sanitization or validation. If an attacker can control the `dir` parameter, they could potentially create directories in unintended locations.)
2. (source, Files.createDirectory, The `source` variable is used to create a relative path without any sanitization or validation. If an attacker can control the `source` variable, they could potentially manipulate the path creation process.)
3. (destDir, Files.createDirectory, The `destDir` variable is used to create a new directory without any sanitization or validation. If an attacker can control the `destDir` variable, they could potentially create directories in unintended locations.)

E. Vulnerability analysis verdict: $$ vulnerability: YES | vulnerability type: CWE-22 | vulnerability name: Improper Limitation of a Pathname to a Restricted Directory ('Path Traversal') | explanation: @\colorbox{yellow}{The code does not validate or sanitize the}@ @\colorbox{yellow}{`dir`, `source`, and `destDir` variables before}@ @\colorbox{yellow}{using them to create a new directory}@. This could allow an attacker to manipulate the path and create directories in unintended locations, leading to a path traversal vulnerability. $$
\end{lstlisting}
\end{figure}

\begin{lstlisting}[style=java,caption={GPT-4 with \df prompt on a real-world Integer Overflow vulnerability},label=fig:cvefixesjava2389,numbers=none,framesep=2pt,frame=lrtb,framerule=1pt,escapechar=@]
// TARGET CWE: CWE-190 (Integer Overflow or Wraparound)

// CODE SNIPPET
public void write(byte[] b, int offset, int length) throws IOException {
    if (b == null) {
      throw new NullPointerException();
    }
    
    if (offset < 0 || offset + length > b.length) {
      throw new ArrayIndexOutOfBoundsException();
    }
    
    write(fd, b, offset, length);
}

// RESPONSE WITH DATAFLOW ANALYSIS INSPIRED PROMPT
Here is a data flow analysis of the given code snippet:
A. Sources: 
1. The byte array 'b'
2. The integer 'offset'
3. The integer 'length'

B. Sinks:
1. The 'write' function

C. Sanitizers:
1. The condition 'if (offset < 0 || offset + length > b.length)'

D. Unsanitized Data Flows:
1. (offset, write, The offset value is directly used in the write function without any sanitization)
2. (length, write, The length value is directly used in the write function without any sanitization)

E. Vulnerability analysis verdict: $$ vulnerability: YES | vulnerability type: CWE-190 | vulnerability name: Integer Overflow or Wraparound | explanation: The code does not properly handle the case where 'offset + length' could overflow. @\colorbox{yellow}{If 'offset' and 'length' are both very large,}@ @\colorbox{yellow}{their sum could wrap around to a negative value}@, passing the bounds check and leading to an ArrayIndexOutOfBoundsException at runtime. $$'
\end{lstlisting}

\begin{lstlisting}[style=java,caption={GPT-4 with \df prompt on a real-world Out of Bounds vulnerability},label=fig:cvefixesjava2241,numbers=none,framesep=2pt,frame=lrtb,framerule=1pt,escapechar=@]
// TARGET CWE: CWE-125 (Out-of-bounds Read)

// CODE SNIPPET
public int encryptWithAd(byte[] ad, byte[] plaintext, int plaintextOffset,
			byte[] ciphertext, int ciphertextOffset, int length)
			throws ShortBufferException {
		int space;
		if (ciphertextOffset > ciphertext.length)
			space = 0;
		else
			space = ciphertext.length - ciphertextOffset;
		if (!haskey) {
					if (length > space)
				throw new ShortBufferException();
			if (plaintext != ciphertext || plaintextOffset != ciphertextOffset)
				System.arraycopy(plaintext, plaintextOffset, ciphertext, ciphertextOffset, length);
			return length;
		}
		if (space < 16 || length > (space - 16))
			throw new ShortBufferException();
		setup(ad);
		encryptCTR(plaintext, plaintextOffset, ciphertext, ciphertextOffset, length);
		ghash.update(ciphertext, ciphertextOffset, length);
		ghash.pad(ad != null ? ad.length : 0, length);
		ghash.finish(ciphertext, ciphertextOffset + length, 16);
		for (int index = 0; index < 16; ++index)
			ciphertext[ciphertextOffset + length + index] ^= hashKey[index];
		return length + 16;
	}

// RESPONSE WITH DATAFLOW ANALYSIS INSPIRED PROMPT
Here is a data flow analysis of the given code snippet:
A. Sources: 
1. The `plaintext` array
2. The `ciphertext` array
3. The `ad` array
4. The `plaintextOffset` integer
5. The `ciphertextOffset` integer
6. The `length` integer

B. Sinks:
1. The `System.arraycopy` method
2. The `encryptCTR` method
3. The `ghash.update` method
4. The `ghash.pad` method
5. The `ghash.finish` method
6. The `ciphertext` array

C. Sanitizers:
1. The `if (ciphertextOffset > ciphertext.length)` check
2. The `if (length > space)` check
3. The `if (space < 16 || length > (space - 16))` check

D. Unsanitized Data Flows:
1. (plaintext, System.arraycopy, if plaintext is larger than the destination array, it could lead to an out-of-bounds read)
2. (ciphertext, ghash.finish, if ciphertextOffset + length + 16 is larger than the size of ciphertext, it could lead to an out-of-bounds read)

E. Vulnerability analysis verdict: vulnerability: YES | vulnerability type: CWE-125 | vulnerability name: Out-of-bounds Read | explanation: @\colorbox{yellow}{The code does not properly check}@ @\colorbox{yellow}{the boundaries of the `ciphertext` array before}@ @\colorbox{yellow}{reading}@ from it, which could lead to an out-of-bounds read if `ciphertextOffset + length + 16` is larger than the size of the `ciphertext` array. 
\end{lstlisting}

\subsection{CWE-125 (Out-of-bounds Read in Java vs. C++)}
\label{appendix:oob_java_cpp}

Listing~\ref{fig:cvefixesjava2241} presents an example of Out-of-bounds read vulnerability in \cvej which GPT-4 predicts correctly.
Codellama-34B also correctly predicts that this snippet is vulnerable.
Listing~\ref{fig:cvefixescpp12915} presents an example from \cvec where Out-of-bounds read vulnerability is not possible but
this is not evident from context. Codellama-34B's explanation seems plausible given that little is known about the pointers in question.

\begin{lstlisting}[style=cpp,caption={Codellama-34B with \df prompt on a real-world Out of Bounds vulnerability (incorrect)},label=fig:cvefixescpp12915,numbers=none,framesep=2pt,frame=lrtb,framerule=1pt]
// CODE SNIPPET NOT VULNERABLE TO CWE-125 (Out-of-bounds Read)
// The pointers cannot be reasoned about with this snippet alone

snmp_ber_decode_length(snmp_packet_t *snmp_packet, uint8_t *length)
{
  if(snmp_packet->used == 0) {
    return 0;
  }

  *length = *snmp_packet->in++;
  snmp_packet->used--;

  return 1;
}

// CODELLAMA-34B WITH THE DATAFLOW ANALYSIS BASED PROMPT

Here is a data flow analysis of the given code snippet:

A. Sources:

1. snmp_packet->in
2. snmp_packet->used

B. Sinks:

1. *length
2. snmp_packet->used

C. Sanitizers:

1. None

D. Unsanitized Data Flows:

1. (snmp_packet->in, *length, This flow could be vulnerable to Out-of-bounds Read (CWE-125) if the value of snmp_packet->in is not properly validated and could be pointing to an invalid memory location.)
2. (snmp_packet->used, snmp_packet->used--, This flow could be vulnerable to Out-of-bounds Read (CWE-125) if the value of snmp_packet->used is not properly validated and could be pointing to an invalid memory location.)

E. Vulnerability analysis verdict: $$ vulnerability: YES | vulnerability type: CWE-125 | vulnerability name: Out-of-bounds Read | explanation: The code snippet does not properly validate the input values, which could lead to an Out-of-bounds Read vulnerability. $$

\end{lstlisting}

\end{document}